\documentclass[aps,prd,floatfix,showpacs,10pt,nofootinbib,twocolumn]{revtex4-1}

\usepackage{amsmath,amssymb,amsfonts,amsthm} 
\usepackage{color}
\usepackage{float}
\usepackage{graphicx}
\usepackage{pstricks,pstricks-add}
\usepackage{pst-plot}
\usepackage[scriptsize,nooneline,hang]{caption}
\usepackage[hang,nooneline,scriptsize]{subfigure}

\definecolor{dark-green}{rgb}{0,0.7,0}
\definecolor{dark-blue}{rgb}{0,0.2,0.5}
\definecolor{med-blue}{rgb}{0,0.7,1}
\definecolor{mblue}{rgb}{0,0.2,1}
\definecolor{cnc}{rgb}{0.8,0,0}
\definecolor{light-red}{rgb}{1,0.8,0.8}
\definecolor{dark-yellow}{rgb}{1,0.8,0}
\definecolor{light-blue}{rgb}{0.8,0.9,1}
\definecolor{grey}{rgb}{0.211,0.211,0.211}
\definecolor{verylight-blue}{rgb}{0.93,0.95,1}
\definecolor{light-yellow}{rgb}{1,0.9,0.8}

\newcommand{\weglassen}[1]{}

\begin{document}

\title{Geodesic motion in the space-time of cosmic strings interacting via magnetic fields}

\author{Betti Hartmann $^{(a)}$ }
\email{b.hartmann@jacobs-university.de}

\author{Valeria Kagramanova $^{(b)}$}
\email{va.kagramanova@uni-oldenburg.de}

\affiliation{
$(a)$ School of Engineering and Science, Jacobs University Bremen, 28759 Bremen, Germany\\
$(b)$ Institut f\"ur Physik, Universit\"at Oldenburg, 26111 Oldenburg, Germany}
\date{\today}

\begin{abstract}
We study the geodesic motion of test particles in the space-time of two Abelian-Higgs strings interacting
via their magnetic fields.
These bound states of cosmic strings constitute a field theoretical realization of p-q-strings which are predicted
by inflationary models rooted in String Theory, e.g. brane inflation. In contrast to
previously studied models describing p-q-strings our model possesses a Bogomolnyi-Prasad-Sommerfield (BPS)
limit. If cosmic strings exist it would be exciting to detect them by direct observation. 
We propose that this can be done by the observation of test particle motion in the
space-time of these objects. In order to be able to make predictions we have to solve the field equations
describing the configuration as well as the geodesic equation
numerically. The geodesics can then be classified according to the test particle's energy, angular
momentum and momentum along the string axis. We find that the interaction of two Abelian-Higgs strings
can lead to the existence of bound orbits that would be absent without the interaction. We also discuss the minimal and maximal radius of orbits and comment on possible
applications in the context of gravitational wave emission.
\end{abstract}

\maketitle

\section{Introduction}

Cosmic strings are topological defects that are predicted to have formed via the Kibble mechanism
\cite{kibble} during one of the phase transitions in the early universe and in the field theoretical
description \cite{no}  can be considered to be an example of a topological soliton. Due to the fact that these
objects can be extremely heavy they were believed to be 
a possible source of the density 
perturbations that led to structure formation and the anisotropies in the 
cosmic microwave background (CMB) \cite{vs}. However, the detailed measurement of the 
CMB power spectrum
as obtained by COBE, BOOMERanG and WMAP showed
that cosmic strings cannot be the main source for these anisotropies. 
However, in recent years it has been suggested that cosmic strings should generically form at the
end of inflation in inflationary models resulting from String Theory \cite{polchinski}
such as brane inflation \cite{braneinflation}. Moreover, cosmic strings seem to be a generic prediction of supersymmetric 
hybrid inflation \cite{lyth} and grand unified based inflationary models \cite{jeannerot}. 
Even though the origin of these cosmic superstrings is String theory, 
their properties can be investigated in the framework of field theoretical models
\cite{saffin,rajantie,salmi,urrestilla}. 
The influence of gravity on field theoretical cosmic superstrings has been studied in 
\cite{hartmann_urrestilla}. 

The field theoretical
model typically used to study cosmic (super)strings is the Abelian-Higgs model, which was shown to possess string-like solutions \cite{no}.
This model contains a complex scalar field minimally coupled to a $U(1)$ gauge field. 
The symmetry breaking pattern of this model is $U(1)\rightarrow$ 1 and is supposed to be a toy model
for the formation of strings in Grand Unified Theories. The gravitational field of an Abelian-Higgs string was also studied \cite{clv,bl}
and it was shown that next to standard string solutions so-called ``Melvin'' solutions exist that
possess a different asymptotic behaviour of the metric functions.

The interaction of two Abelian-Higgs strings via their magnetic fields
has been studied \cite{ha}. It was shown that bound states can form
and that a new Bogomolnyi-Prasad-Sommerfield (BPS) bound \cite{bps} exists for particular choices of the coupling constants in the model.

Since cosmic (super)strings are a prediction of String Theory and Grand Unified Theories it would be exciting
to detect these objects. There has 
been considerable effort in 
numerically modeling cosmic string networks to obtain
CMB power and polarization spectra \cite{cmb,cmb2}. Comparison with observations has
shown that cosmic strings might well contribute considerably to the
energy density of the universe. 
In this paper, we discuss another possibility to detect cosmic strings
namely through the motion of test bodies in such string space-times. As such light deflection, i.e. the motion
of massless test particles in cosmic string space-times has been used to suggest cosmic string candidates \cite{csl1}.
The test particle motion in different space-times containing
cosmic strings has been investigated in \cite{ag,gm,cb,Ozdemir2003,Ozdemir2004}, while the complete set of
orbits of test particles in the space-time of a black hole pierced by an infinitely thin cosmic string has been given
for a Schwarzschild black hole in \cite{hhls1} and for a Kerr black hole in \cite{hhls2}. Moreover,
the geodesic motion of test particles in field theoretical cosmic string space-times has been given for
Abelian-Higgs strings in \cite{hartmann_sirimachan} and for cosmic superstrings in 
\cite{hartmann_laemmerzahl_sirimachan}.

In this paper we are aiming at studying the motion of test particles in the space-time of two Abelian-Higgs
strings that form bound states by interacting via their magnetic fields. Note that
the form of interaction considered here has been used in the description of
new theoretical models of the dark matter sector \cite{dark}.
The bound states of two Abelian-Higgs strings can be thought of as a field theoretical realization of cosmic superstrings, so-called
p-q-strings. In the original field theoretical models the two strings interact via 
a potential term \cite{saffin,rajantie,salmi,urrestilla}. However, in these
models there are no bound states that fulfill the BPS limit. This is different
in our model, where a BPS state exists for certain values of the
coupling constants \cite{ha}.

Our paper is organised as follows: in Section II we give the field theoretical model describing dark strings
interacting with cosmic strings as well as the geodesic equations describing test particle motion
in the space-time of these strings. In Section III we present our numerical results and we conclude in Section IV.

\section{The Model}
The field theoretical model to describe the interaction of
a dark string with a cosmic string reads \cite{ha}
\begin{eqnarray}
S&=&\int d^4x\sqrt{-g}\left(\frac{1}{16\pi G}R+\mathcal{L}_{\rm m}\right)  \ ,
\label{action}
\end{eqnarray}
where $R$ is the Ricci scalar and $G$ is Newton's constant. 
The matter Lagrangian $\mathcal{L}_{\rm m}$ is given by 
\begin{eqnarray}
\mathcal{L}_{\rm m}&=&D_{\mu}\phi(D^{\mu}\phi)^*-
\frac{1}{4}F_{\mu\nu}F^{\mu\nu}+D_{\mu}\xi(D^{\mu}\xi)^* \nonumber \\ &&  -\frac{1}{4}H_{\mu\nu}H^{\mu\nu}-u(\phi,\xi)
+ \frac{\alpha}{2} F_{\mu\nu}H^{\mu\nu}
\end{eqnarray}
with the covariant derivatives  
$D_{\mu}\phi$ = $\nabla_{\mu}\phi$ - $i e_1 A_{\mu}\phi$, 
$D_{\mu}\xi$ = $\nabla_{\mu}\xi$ - $i e_2 B_{\mu}\xi$ of the two complex scalar 
field
$\phi$ and $\xi$.  The field strength tensors are 
$F_{\mu\nu}=\nabla_{\mu}A_{\nu}-\nabla_{\nu}A_{\mu}=\partial_{\mu}A_{\nu} - \partial_{\nu}A_{\mu}$, 
$H_{\mu\nu}=\nabla_{\mu}B_{\nu} - \nabla_{\nu}B_{\mu}= \partial_{\mu}B_{\nu} - \partial_{\nu}B_{\mu}$ 
of two U(1) gauge potential $A_{\mu}$, $B_{\nu}$ with coupling constants $e_1$ 
and $e_2$. $\nabla_{\mu}$ denotes the gravitational covariant derivative. 
Finally, the potential $V(\phi,\xi)$ reads:
\begin{eqnarray}
u(\phi,\xi)&=&\frac{\lambda_1}{4}(\phi\phi^*-\eta_1^{2})^2+\frac{\lambda_2}{4}(\xi\xi^*
-\eta_2^{2})^2 \ ,
\label{Vpq}
\end{eqnarray}
where $\lambda_1$ and $\lambda_2$ are the self-couplings of the two scalar fields. 
$\eta_1$ and $\eta_2$ are the
vacuum expectation values of the scalar fields. 
The term proportional to $\alpha$ is the interaction term \cite{dark}.
We will assume that $\alpha \geq 0$ hence considering bound states of
cosmic strings.

In the following, we associate the dark string
to the fields $A_{\mu}$ and $\phi$,
while the cosmic strings are described by the fields $B_{\mu}$ and $\xi$.
The Higgs fields have masses $M_{H,i}=\sqrt{2\lambda_i} \eta_i$, 
while the gauge boson masses are $M_{W,i}=e_i \eta_i$, $i=1,2$.
Note that the Lagrangian describes effectively two coupled Abelian--Higgs models.

The most general static cylindrically symmetric line element invariant under boosts along the $z$-direction is
\begin{eqnarray}
\label{metric}
ds^2=N^2dt^2-d\rho^2-L^2d\varphi^2-N^2dz^2  \ , \label{cysymmetric}
\end{eqnarray}
where $N$ and $L$ are functions of $\rho$ only.

The non-vanishing components of the Ricci tensor $R_{\mu}^{\nu}$ then read \cite{clv}:
\begin{eqnarray}
&& R_0^0=-\frac{(LNN')'}{N^2 L} \ \ , \ \ R_{\rho}^{\rho} = -\frac{2N''}{N}-\frac{L''}{L} \ \ , \nonumber \\
&& \ \ R_{\varphi}^{\varphi}= -\frac{(N^2 L')'}{N^2 L} \ \ , \ \ R_z^z=R_0^0
\end{eqnarray}
where the prime denotes the derivative with respect to $\rho$. 

For the matter and gauge fields, we have:
\begin{equation}
\phi(\rho,\varphi)=\eta_1 f(\rho)e^{i n\varphi} \ \ , \ \  \xi(\rho,\varphi)=\eta_1 h(\rho)e^{i m\varphi} \ , 
\end{equation}
 and
\begin{eqnarray}
&& A_{\mu}dx^{\mu}=\frac {1}{e_1}(n-R(\rho)) d\varphi \ \ , \nonumber \\
&& B_{\mu}dx^{\mu}=\frac {1}{e_2}(m-P(\rho)) d\varphi \ .
\end{eqnarray}
$n$ and $m$ are integers indexing the vorticity of the two Higgs fields  around the $z-$axis.
The magnetic fields associated to the solution can be given when noting that the gauge part of the Lagrangian
density can be rewritten as follows \cite{vachaspati}:
\begin{eqnarray}
&& -\frac{1}{4} F_{\mu\nu} F^{\mu\nu}-\frac{1}{4} H_{\mu\nu} H^{\mu\nu}+ 
\frac{\alpha}{2} F_{\mu\nu}H^{\mu\nu} \nonumber \\
&&  \Rightarrow \ \  -\frac{1}{4} \tilde{F}_{\mu\nu}
\tilde{F}^{\mu\nu} -\frac{1}{4}(1-\alpha^2) H_{\mu\nu} H^{\mu\nu}
\end{eqnarray}
with $\tilde{F}_{\mu\nu}=\partial_{\mu} \tilde{A}_{\nu}- \partial_{\nu} \tilde{A}_{\mu}$
where $\tilde{A}_{\mu}=A_{\mu}-\alpha a_{\mu}$.

The magnetic fields associated to the fields $\tilde{A}_{\mu}$ and $a_{\mu}$ have only a component in
$z$-direction. These components read: 
\begin{eqnarray}
\label{magnetic}
&& \tilde{B}_z(\rho)=\frac{-P'(\rho)+\frac{\alpha}{g} R'(\rho)}{e_1 L(\rho)} \ \ \ 
{\rm and} \nonumber \\  && b_z(\rho)=-\sqrt{1-\alpha^2}\frac{R'(\rho)}{e_2 L(\rho)}  \ ,  
\end{eqnarray}
respectively. The corresponding magnetic fluxes $\int d^2x \ B$ are
\begin{equation}
 \tilde{\Phi}= \frac{2\pi}{e_1}\left(n-\frac{\alpha}{g} m\right) \ \ {\rm and} \ \ 
\varphi=\sqrt{1-\alpha^2} \ \frac{2\pi m}{e_2} \ ,
\end{equation}
respectively. Obviously, these magnetic fluxes are not quantized for generic $\alpha$
and the two strings interact via their magnetic fields.

Finally, the deficit angle $\delta=8\pi G\mu$ of the solution can be read off directly from the derivative of the
metric function $L(\rho)$. For string-like solutions, the metric functions behave like 
$N(\rho\rightarrow \infty) \rightarrow c_1$ and $L(\rho\rightarrow \infty) 
\rightarrow c_2 \rho + c_3$, where $c_1$, $c_2$ and $c_3$ are constants.
The deficit angle is then given by:
\begin{equation}
\delta=2\pi (1-L'\vert_{\rho=\infty})=2\pi(1-c_2)  \ .
\end{equation}

\subsubsection{Equations of motion and boundary conditions}
We define the following dimensionless quantities
\begin{equation}
\label{scaling}
\rho\rightarrow \frac{\rho}{e_1\eta_1} \ \ \ , \ \ \ L\rightarrow \frac{L}{e_1\eta_1}\ ,
\end{equation}
such that $\rho$ now measures the radial
distance in units of $M_{W,1}/\sqrt{2}$. 

Then, the total Lagrangian ${\cal L}_m \rightarrow {\cal L}_m/(\eta_1^4 e_1^2)$ depends only on the following dimensionless coupling constants

\begin{eqnarray}
&& \gamma=8\pi G\eta_1^2 \ \ ,  \ \ g=\frac{e_2}{e_1} \ \ , \  \ \
q=\frac{\eta_2}{\eta_1} \ , \nonumber \\ &&  \beta_i=\frac{\lambda_i}{e_1^2}=\frac{M^2_{H,i}}{M^2_{W,1}}\frac{\eta_1^2}{\eta_i^2} \ , \ i=1,2  \, . 
\end{eqnarray}
 Varying the action with respect to the matter fields we
obtain a system of five non-linear differential equations. The Euler-Lagrange equations for the matter field functions read:
\begin{equation}
\label{eq1}
\frac{(N^2Lf')'}{N^2L}=\frac{R^2 f}{L^2}+\frac{\beta_1}{2} f(f^2-1) \ , 
\end{equation}
\begin{equation}
\label{eq2}
\frac{(N^2 L h')'}{N^2 L}=\frac{P^2 h}{L^2}+\frac{\beta_2}{2} (h^2-q^2)h \ ,
\end{equation}

\begin{equation}
\label{eq4}
(1-\alpha^2)\frac{L}{N^2}\left(\frac{N^2P'}{L}\right)'=2 g^2 h^2 P+ 2\alpha g R f^2 \ ,
\end{equation}
\begin{equation}
\label{eq5}
(1-\alpha^2)\frac{L}{N^2}\left(\frac{N^2R'}{L}\right)'=2 f^2 R + 2\alpha g P h^2 \ ,
\end{equation}
while the Einstein equations are
\begin{eqnarray}
\label{eq6}
\frac{(LNN')'}{N^2 L}&=& \gamma\left[\frac{(P')^2}
{2 g^2 L^2}+ \frac{(R')^2}{2 L^2} -\frac{\alpha}{g}\frac{R'P'}{L^2}
-u(f,h)\right] \qquad
\end{eqnarray}
and:
\begin{eqnarray}
\label{eq7}
\frac{(N^2L')'}{N^2L}&=&-\gamma\left[\frac{2 h^2 P^2}
{L^2}+   \frac{2 R^2 f^2}{L^2}+\frac{(P')^2}{2 g^2 L^2}  \right. \nonumber \\ 
&& \left. + \frac{(R')^2}{2 L^2 } -\frac{\alpha}{g}\frac{R'P'}{L^2} + u(f,h)\right]  \ ,
\end{eqnarray}
where the prime now and in the following denotes the derivative with respect to $\rho$. The potential
now reads:
\begin{equation}
\label{potential_scaled}
 u(f,h)=\frac{\beta_1}{4}\left(f^2-1\right)^2 + \frac{\beta_2}{4}\left(h^2-q^2\right)^2   \ .
\end{equation}

These equations have to be solved numerically subject to appropriate boundary conditions.
We require the solution to be regular at $\rho=0$ which implies
\begin{equation}
f(0)=0 \ , \  h(0)=0 \  , \ P(0)=m \ , \ R(0)=n
\label{bc1}
\end{equation}
for the matter fields and 
\begin{equation}
\label{zero}
N(0)=1, \ N'(0)=0, \ L(0)=0 \ , \ L'(0)=1 \,
\end{equation}
for the metric fields. 
The finiteness of the energy per unit length requires:
\begin{equation}
f(\infty)=1, \ h(\infty)=q   \ ,  \ P(\infty)=0 \ , \ R(\infty)=0  \ .
\end{equation}
A BPS limit
exists \cite{ha} if $f\equiv h$ ($q=1$) and $P=R$ ($n=m$, $g=1$) for
\begin{equation}
\beta_1=\beta_2\equiv \beta=\frac{2}{1-\alpha}  \ .
\end{equation}
In this limit the metric function $N(\rho)\equiv 1$, while the remaining functions fulfill
the BPS equations
\begin{equation}
 f'=\frac{Pf}{L} \ \ , \ \ (1-\alpha) \frac{P'}{L}=f^2-1 \ ,
\end{equation}
\begin{equation}
\frac{L''}{L}=-2\gamma\left(\frac{2P^2f^2}{L^2} + \frac{1}{1-\alpha}(f^2-1)^2\right)  \ .
\end{equation}

\subsection{The geodesic equation}
The Lagrangian $\mathcal{L}_g$ describing geodesic motion of a test particle 
in the static cylindrically symmetric space-time (\ref{cysymmetric}) reads
\begin{eqnarray}
\label{geolag}
&& 2 \mathcal{L}_{\rm g}=g_{\mu\nu}\frac{dx^{\mu}}{d\tau}\frac{dx^{\nu}}{d\tau}=\varepsilon  \nonumber \\
&& =  N^2\left(\frac{dt}{d\tau}\right)^2 -\left(\frac{d\rho}{d\tau}\right)^2
-L^2\left(\frac{d\varphi}{d\tau}\right)^2-N^2\left(\frac{dz}{d\tau}\right)^2 , \qquad
\label{lag}
\end{eqnarray}
where $\varepsilon = 0, 1$ for massless or massive test particles, respectively 
and $\tau$ is an affine parameter that corresponds to the proper time for massive test particles
moving on time-like geodesics. The space-time has three Killing vectors 
$\frac{\partial}{\partial t}$, $\frac{\partial}{\partial \varphi}$ and
$\frac{\partial}{\partial z}$ which lead to the following constants of motion: the energy $E$, 
the angular momentum $L_z$ along the string axis ($z$-axis) 
and the momentum $P_z$
\begin{eqnarray}
N(\rho)^2\frac{dt}{d\tau}=:E \ , \ \ 
L(\rho)^2\frac{d\varphi}{d\tau}=:L_z \ , \ \ 
N(\rho)^2\frac{dz}{d\tau}=:P_z \ . \qquad
\end{eqnarray}
Using the rescaling (\ref{scaling}) the constants of motion must be rescaled 
according to $E$ $\rightarrow$ $E/(e_1\eta_1)$, 
$P_z$ $\rightarrow$ $P_z/(e_1\eta_1)$, $L_z$ $\rightarrow$ $L_z/(e_1\eta_1)^2$.  
We then find from (\ref{lag})
\begin{eqnarray}
\varepsilon &=& N^2\left(\frac{dt}{d\tau}\right)^2-\left(\frac{d\rho}{d\tau}\right)^2
-L^2\left(\frac{d\varphi}{d\tau}\right)^2-N^2\left(\frac{d z}{d\tau}\right)^2 \nonumber \\
&=&\frac{E^2-P_z^2}{N^2}-\left(\frac{d\rho}{d\tau}\right)^2-\frac{L_z^2}{L^2}  \label{udot}  \ .
\end{eqnarray}
Using the constants of motion we find from (\ref{geolag})
\begin{eqnarray}
\left(\frac{d\rho}{d\tau}\right)^2 &=&
\frac{1}{N^2}
\left[E^2 - N^2\left( \varepsilon + \frac{P_z^2}{N^2}+\frac{L_z^2}{L^2} \right) \right]\ . \label{Nrdotsqr}
\end{eqnarray}
The left hand side of (\ref{Nrdotsqr}) is always positive. Following \cite{kkl} we can then rewrite this equation as
\begin{eqnarray}
\label{geo1}
\left(\frac{d\rho}{d\tau}\right)^2 &=& \frac{1}{N^2}
\left[{\cal E} -V_{\rm eff}(\rho) \right] \label{afromstring}  \ ,
\end{eqnarray}
where
\begin{eqnarray}
\label{effective_pot}
V_{\rm eff}(\rho)&=&N^2\left( \varepsilon + \frac{P_z^2}{N^2}+\frac{L_z^2}{L^2} \right) \label{Veff}
\end{eqnarray}
is the effective potential and ${\cal E}=E^2$.

In the following, we would like to find $t(\rho)$, $\varphi(\rho)$ and $z(\rho)$. For this,
we rewrite the geodesic equation in the form
\begin{eqnarray}
d\varphi&=&\pm \frac{L_z N d\rho}{L(\rho)^2\left( E^2 - N^2 \left( \varepsilon + \frac{P_z^2}{N^2}+\frac{L_z^2}{L^2} \right) \right)^{1/2}}\label{phiint}  \ , \\
dz&=&\pm \frac{P_z d\rho}{N(\rho) \left( E^2 - N^2 \left( \varepsilon + \frac{P_z^2}{N^2}+\frac{L_z^2}{L^2} \right) \right)^{1/2}}\ , \label{zint}\\
dt&=&\pm \frac{E d\rho}{N(\rho) \left( E^2 - N^2 \left( \varepsilon + \frac{P_z^2}{N^2}+\frac{L_z^2}{L^2} \right) \right)^{1/2}}\label{tint}  \ .
\end{eqnarray}

The solution for each component can then be calculated as a 
function of $\rho$ by using numerical integration methods.

\section{Numerical results}
The solutions to the equations (\ref{eq1})-(\ref{eq7}) are only known numerically. We have solved these equations
using the ODE solver COLSYS \cite{colsys}. The solutions have relative errors on the order of $10^{-9}-10^{-13}$.
The solution corresponds to two Abelian-Higgs strings
interacting via their magnetic fields in curved space-time and has been studied in detail in
\cite{ha}. In the following we will set $g=q=1$ unless otherwise stated. The equations of motion are integrated
numerically in Fortran using the Fortran Subroutines for
Mathematical Applications--IMSL MATH/LIBRARY. The accuracy of the integration as estimated from application of 
the method for the integration of geodesics in Schwarzschild space-time is on the order of $10^{-5}$.

\subsection{The effective potential}
It is clear from (\ref{geo1}) and (\ref{effective_pot}) that we need
to require $E^2 > V_{\rm eff}$ in order to find orbits. In addition the values
of $\rho$ for which $E^2=V_{\rm eff}$ correspond to the turning points of the motion.
For $L_z\neq 0$ the effective potential tends to infinity for $\rho\rightarrow 0$. Physically
this corresponds to an infinite potential barrier resulting from the
angular momentum of the particle. As such the test particle can never reach the
string axis $\rho=0$ for $L_z\neq 0$. For $L_z=0$ on the other hand, the potential
has a finite value at $\rho=0$ and if $E^2$ is greater than this value the particles
can reach the string axis. Asymptotically the potential tends to $V_{\rm eff}(\rho\rightarrow \infty)\rightarrow
c_1^2\varepsilon + P_z^2$.

\subsubsection{Massless test particles}
In \cite{gibbons} it was shown that 
for a general cosmic string space--time with topology $\mathbb{R}^2\times \Sigma$
massless test particles must move on geodesics that escape to infinity in both directions, i.e.
bound orbits of massless test particles are not possible. 
The assumption made in \cite{gibbons} is that $\Sigma$ must have positive Gaussian curvature.
In \cite{hartmann_sirimachan} and \cite{hartmann_laemmerzahl_sirimachan} it was shown
that in the space-time of Abelian-Higgs strings and cosmic superstrings, respectively, this
is even true if $\Sigma$ has negative Gaussian curvature close to the string axis.
In the Appendix we give a general proof that the effective potential cannot have 
local extrema in this case and hence bound orbits are not possible.
We don't present any plots of the effective potential here since it is simply monotonically decreasing (for $L_z\neq 0$)
or constant (for $L_z=0$). 

\subsubsection{Massive test particles}
In contrast to massless test particles we expect massive test particles to be able to move on bound orbits
around the cosmic strings. That this is possible for {\it finite width} cosmic strings was shown
in \cite{hartmann_sirimachan,hartmann_laemmerzahl_sirimachan}. We will demonstrate in the following
that this is also possible in our model.

Bound orbits possess two turning points and we need hence to require that the
potential should not be monotonically decreasing on the full $\rho$ interval, but
have local extrema. Remembering that $L' > 0$ it is obvious from the form of the
potential that solutions to $dV_{\rm eff}/d\rho=0$ exist only if $N' > 0$. 

\begin{figure*}[t]
\begin{center}
\subfigure[][$L_z^2=0.0, P_z^2=4.0$]{\label{c1_pot1}\includegraphics[width=7.0cm]{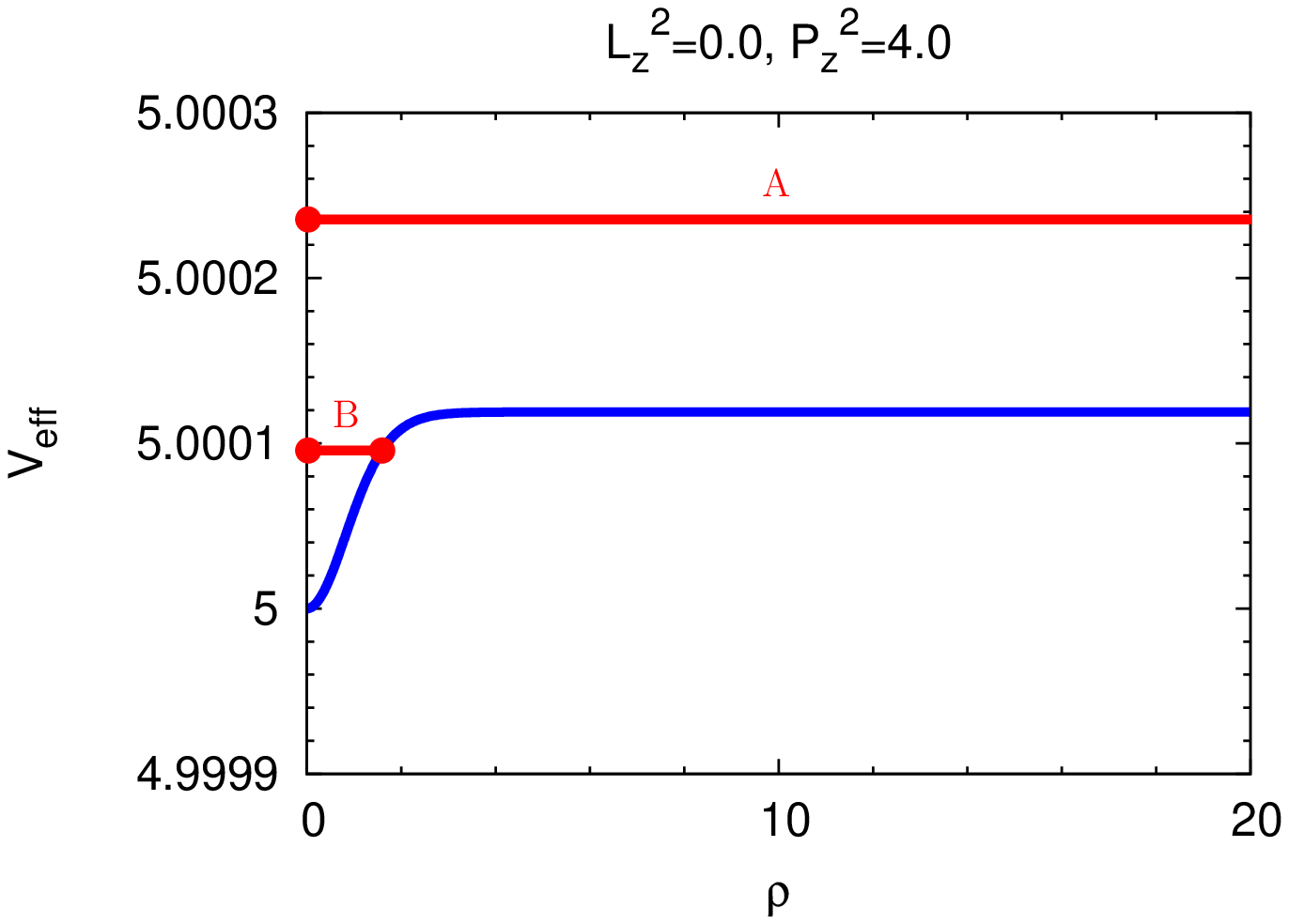}}
\subfigure[][$L_z^2=1.0\cdot 10^{-5}, P_z^2=4.0$]{\label{c1_pot12}\includegraphics[width=7.0cm]{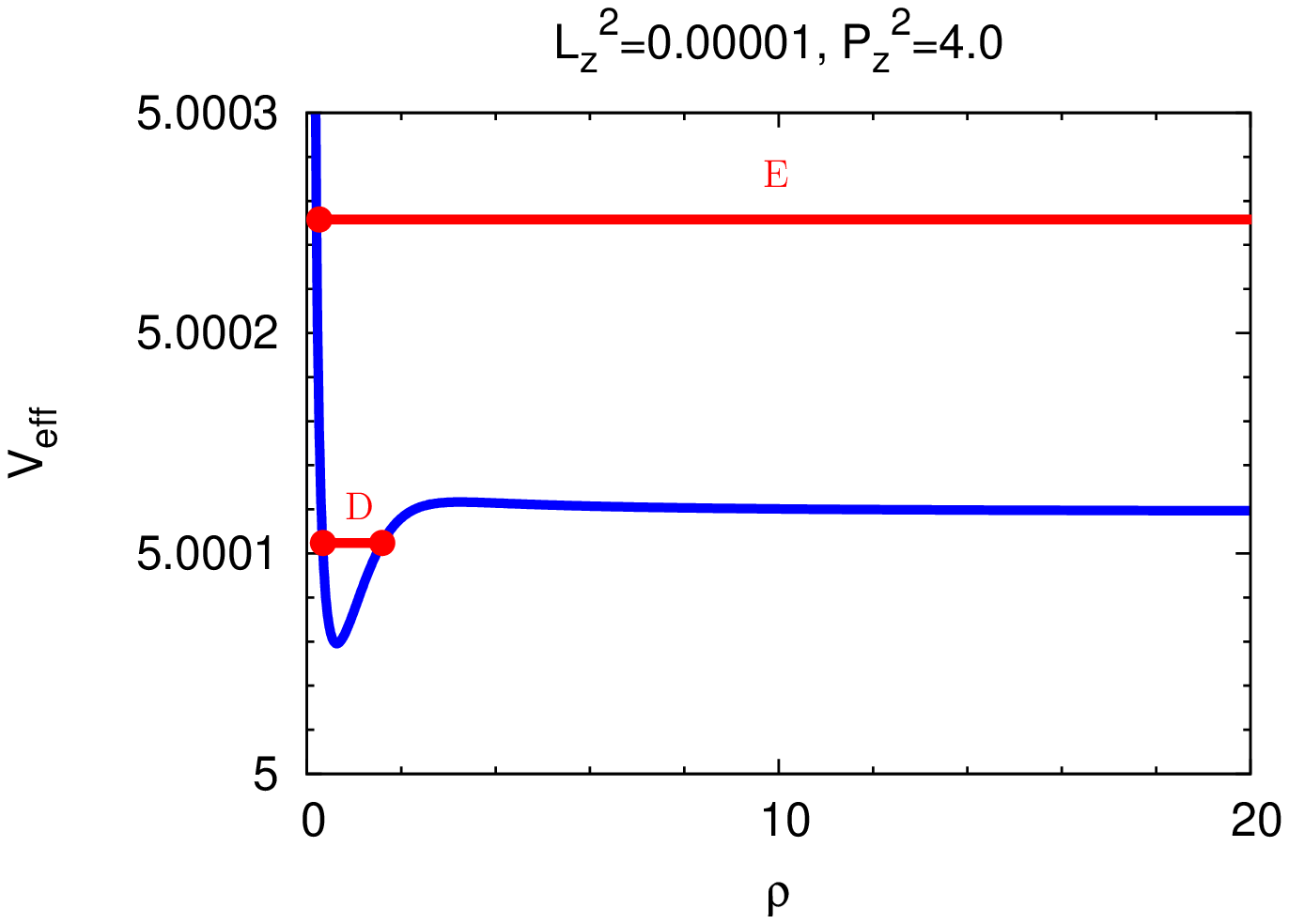}}\\
\subfigure[][$L_z^2=5.0\cdot 10^{-5}, P_z^2=4.0$]{\label{c1_pot2}\includegraphics[width=7.0cm]{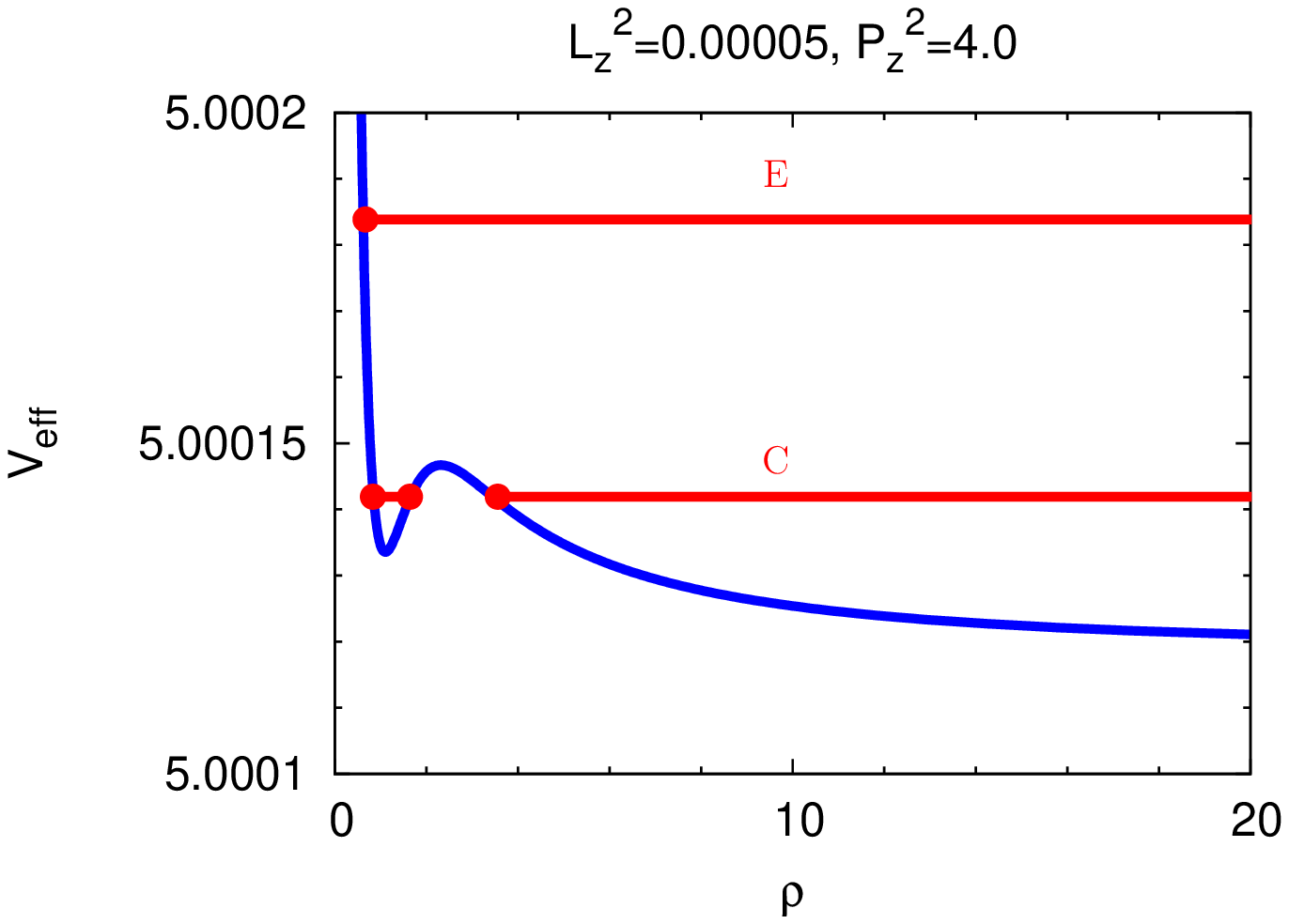}}
\subfigure[][$L_z^2=5.0\cdot 10^{-4}, P_z^2=4.0$]{\label{c1_pot3}\includegraphics[width=7.0cm]{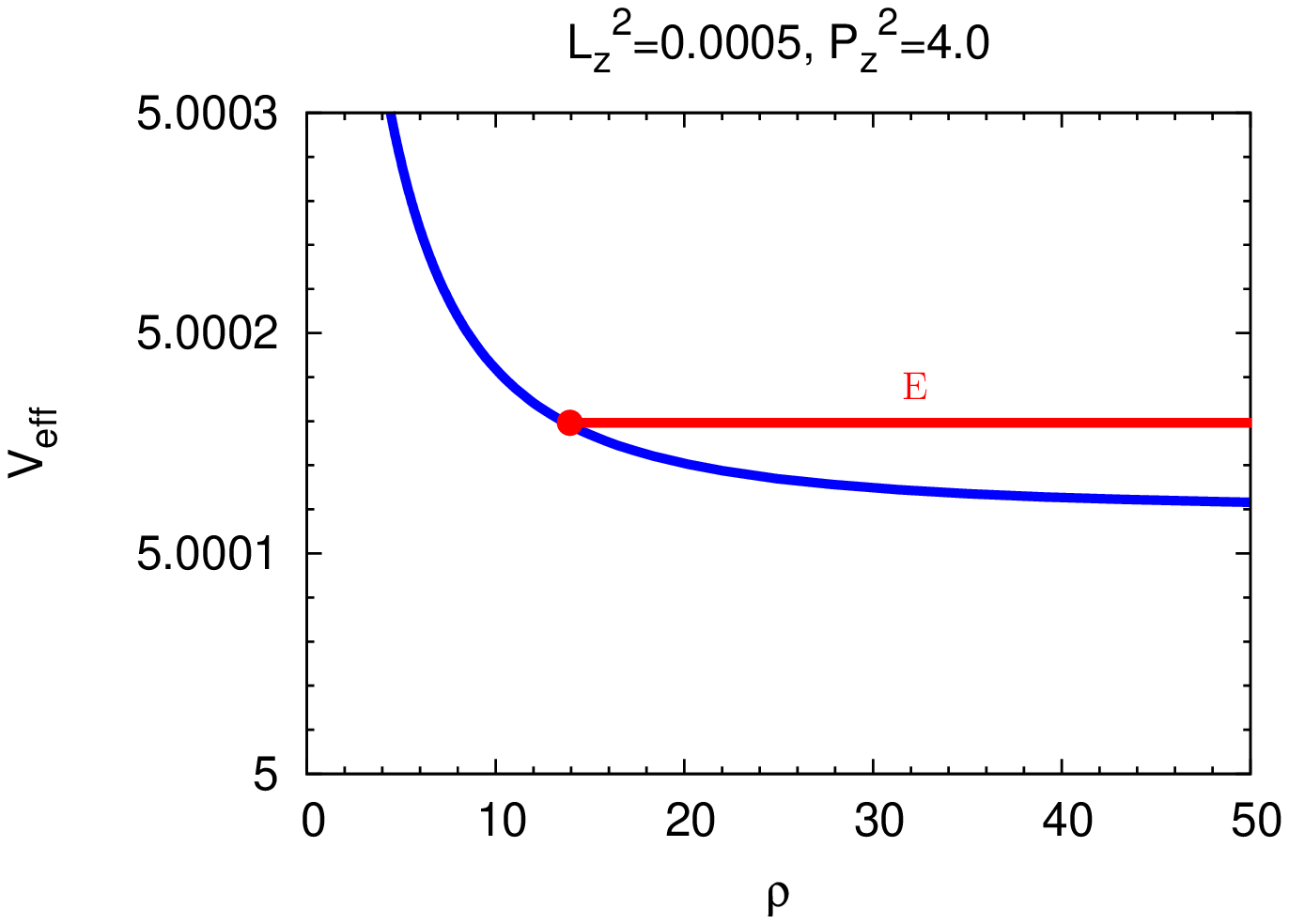}}
\end{center}
\caption{We show the effective potential for 
$n=m=1, \beta_1=\beta_2=2.0, \alpha=0.001, g=1.0, q=1.0$, $P_z^2=4$ and different
values of the angular momentum $L_z$\label{c1pots}. The red horizontal
lines $A$-$E$ indicate the type of orbits possible. See also Table~\ref{TypesOfOrbits1}.}
\end{figure*}

For $\alpha=0$ the two Abelian-Higgs strings do not interact directly.
As mentioned above, geodesic motion in the space-time of an 
Abelian-Higgs string has been studied in \cite{hartmann_sirimachan}
and it was found that bound orbits of massive particles are only possible if the 
Higgs boson mass
is smaller than the gauge boson mass or equivalently if the scalar core of the string
is larger than the core of the corresponding flux tube since in that
case $N' > 0$. If the Higgs boson mass is equal (larger) than the gauge boson
mass it was found that $N'=0$ ($N' < 0$). Hence the effective potential does not
have local extrema and bound orbits are not possible.
This changes if one couples the two Abelian-Higgs models
via a potential interaction term \cite{hartmann_laemmerzahl_sirimachan}. 
In this present paper, the two sectors
interact via the magnetic fields of the cosmic strings. 
In analogy to \cite{hartmann_laemmerzahl_sirimachan} we are hence first interested to
see whether bound orbits exist also when the radii of the flux and scalar cores are equal, 
i.e. for $\beta_1=\beta_2=2$.
In Fig.\ref{c1pots} we show the effective potential for massive test particles ($\varepsilon=1$) with $n=m=1$, 
$\beta_1=\beta_2=2.0$ and $\alpha=0.001$ for different
values of the angular momentum $L_z$ with linear momentum $P_z=4.0$. 
For vanishing angular momentum (see Fig.\ref{c1_pot1}) we find
that the angular momentum barrier at $\rho=0$ disappears and the particle moves from 
infinity to the string axis on a terminating escape orbit 
if ${\cal E}$ is larger than the asymptotic value of the effective potential (A) or on
a terminating orbit if ${\cal E}$ is smaller than the asymptotic value (B), respectively
(see also Table~\ref{TypesOfOrbits1} for the
classification of these orbits). For non-vanishing angular 
momentum we find that for some range of $L_z$ bound orbits exist.
This can be seen in Fig.\ref{c1_pot12} for $L_z^2=1\cdot 10^{-5}$,
where for ${\cal E}$ smaller than the asymptotic value of the effective potential
only a bound orbit (D) exists, while for ${\cal E}$ larger than the asymptotic value only
an escape orbit (E) is present. 
Increasing $L_z$ further (Fig.\ref{c1_pot2} for $L_z=5\cdot 10^{-5}$) we find that
an escape orbit and a bound orbit exist (C) for the same value of ${\cal E}$ if this
value is larger than  the asymptotic value of the effective potential, but smaller than the
local maximum. For ${\cal E}$ larger than the local maximum, only an escape orbit is possible (E). 

The effective potential possesses local
extrema such that for ${\cal E}$ larger than  the minimum of the potential and 
smaller than the maximum of the potential there are three turning points and 
bound orbits as well as escape orbits exist. For larger values of $L_z$ (see Fig.\ref{c1_pot3}) 
the local
extrema have disappeared and only 
escape orbits exist. Hence, we conclude that the interaction
between the cosmic strings leads to the existence of bound orbits.  
We summarize possible types of orbits in Table~\ref{TypesOfOrbits1}.

\begin{figure*}[t]
\begin{center}
\subfigure[$L_z^2=10^{-4}, P_z^2=2.5$, $\beta_1=\beta_2=2.1$, $\alpha=0.05$
]{\label{b_1_2_21_pot}\includegraphics[width=7.0cm]{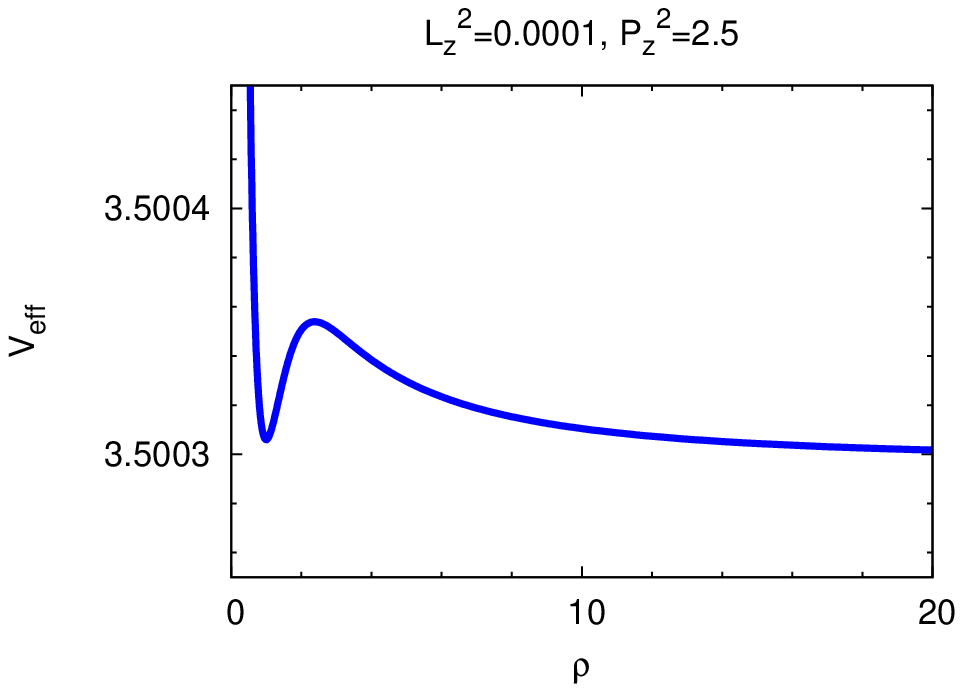}}
\subfigure[$L_z^2=5 \cdot 10^{-4}, P_z^2=4.0$, $\beta_1=\beta_2=0.2$, $\alpha=0.001$]
{\label{c4pot_Lz2=0.0005}\includegraphics[width=7.0cm]{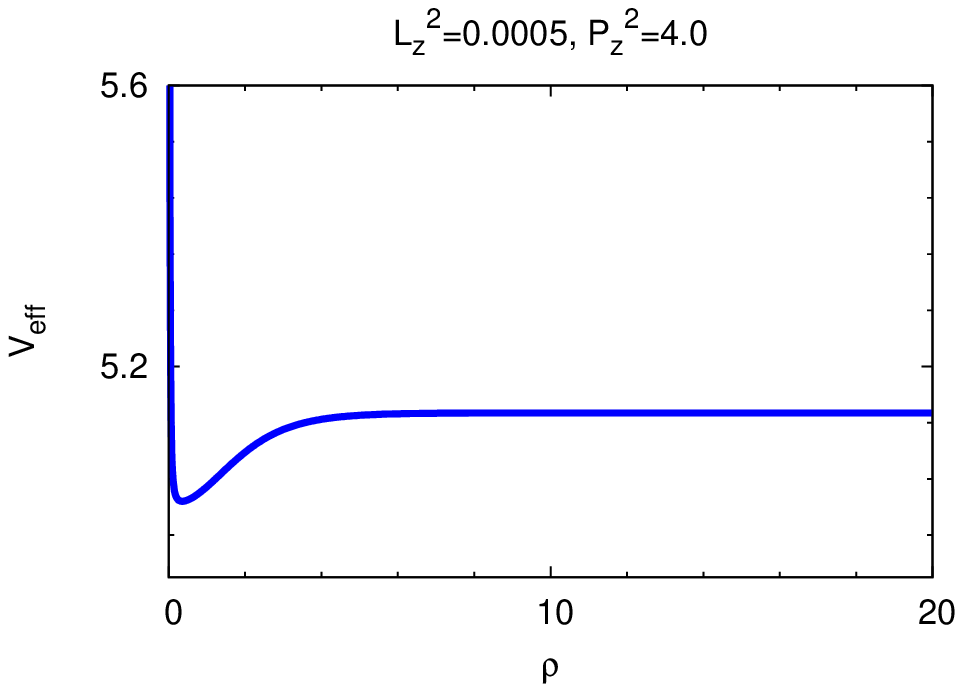}}
\end{center}
\caption{We show the effective potential for $n=m=1, g=1.0, q=1.0$ and different values of $\beta_1=\beta_2$, $\alpha$ and
$P_z$, $L_z$.}
\end{figure*}

\begin{table*}[t]
\begin{center}
\begin{tabular}{|ccll|}\hline
type  & turning points 
& range of $\rho$ & orbit \\ \hline\hline
A  & 0  & 
\begin{pspicture}(-2,-0.2)(3,0.2)
\psline[linewidth=0.5pt]{->}(-2,0)(3,0)

\psline[linewidth=1.2pt](-2,0)(3,0)
\end{pspicture} 
& terminating escape orbit \\  \hline
B  & 1  & 
\begin{pspicture}(-2,-0.2)(3,0.2)
\psline[linewidth=0.5pt]{->}(-2,0)(3,0)
\psline[linewidth=1.2pt]{-*}(-2,0)(0,0)
\end{pspicture} 
 & terminating orbit \\ \hline
C  & 3  & 
\begin{pspicture}(-2,-0.2)(3,0.2)
\psline[linewidth=0.5pt]{->}(-2,0)(3,0)
\psline[linewidth=1.2pt]{*-*}(-0.5,0)(1.5,0)
\psline[linewidth=1.2pt]{*-}(2.0,0)(3,0)
\end{pspicture} 
 & bound and escape orbits \\ \hline
D  & 2  & 
\begin{pspicture}(-2,-0.2)(3,0.2)
\psline[linewidth=0.5pt]{->}(-2,0)(3,0)
\psline[linewidth=1.2pt]{*-*}(-0.5,0)(1.5,0)
\end{pspicture} 
 & bound orbit \\ \hline
E & 1 & 
\begin{pspicture}(-2,-0.2)(3,0.2)
\psline[linewidth=0.5pt]{->}(-2,0)(3,0)
\psline[linewidth=1.2pt]{*-}(0.5,0)(3,0)
\end{pspicture} 
 & escape orbit \\ \hline\hline
\end{tabular}
\caption{Types of orbits possible 
for massive test particles moving in the space-time of cosmic strings interacting via their magnetic
fields   \label{TypesOfOrbits1}}
\end{center}
\end{table*}

Next we have investigated whether bound orbits also exist for $\beta_1$ or $\beta_2$ 
larger than $2$. In Fig.\ref{b_1_2_21_pot} we show the 
potential for $\beta_1=\beta_2=2.1$ and $\alpha=0.05$. Clearly, the potential possesses
local extrema and we can choose the energy such that up to three turning points exist
and hence bound orbits are possible.

\begin{figure*}[t]
\begin{center}
\subfigure[][$L_z^2=0.0, P_z^2=0.5$]{\label{c3_pot1}\includegraphics[width=5.0cm]{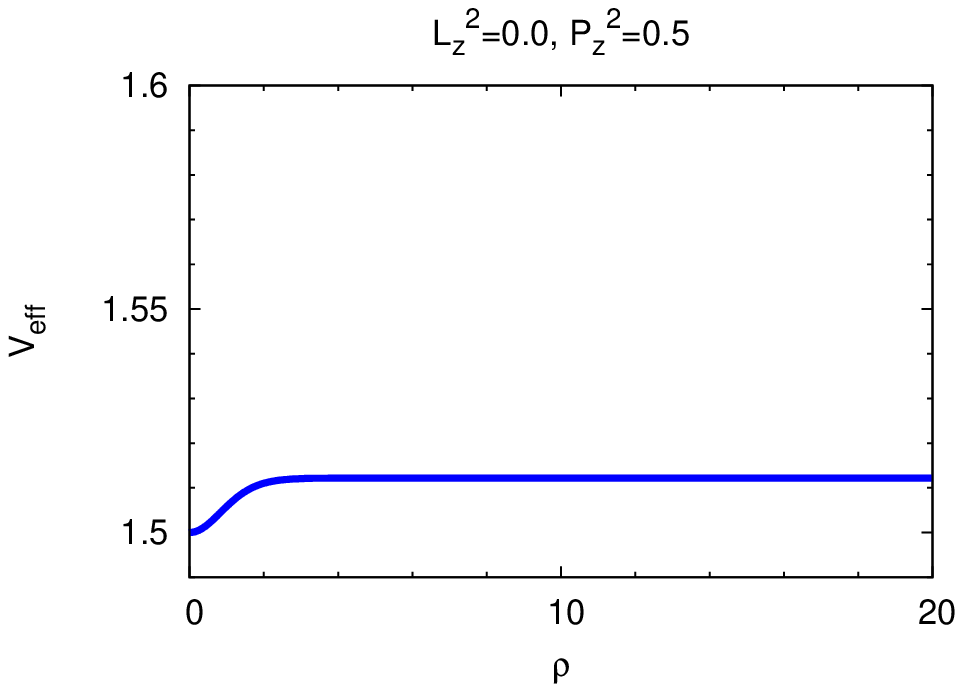}}
\subfigure[][$L_z^4=5.0 \cdot 10^{-5}, P_z^2=0.5$]{\label{c3_pot2}\includegraphics[width=5.0cm]{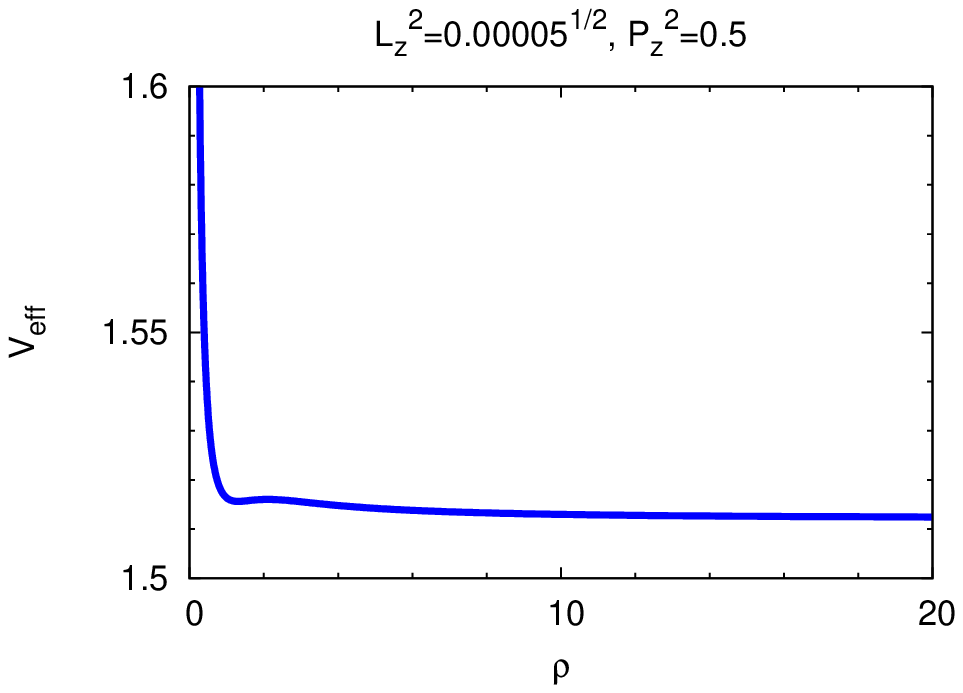}}
\subfigure[][$L_z^2=5.0 \cdot 10^{-5}, P_z^2=0.5$]{\label{c3_pot3}\includegraphics[width=5.0cm]{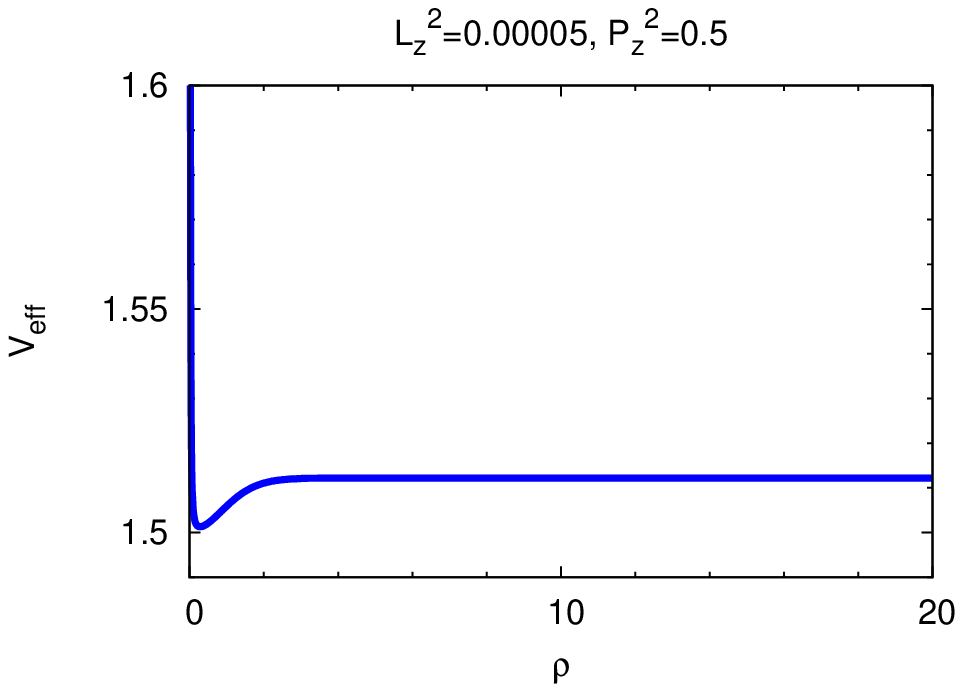}}
\subfigure[][$L_z=0.2, P_z^2=0.5$]{\label{c3_pot4}\includegraphics[width=5.0cm]{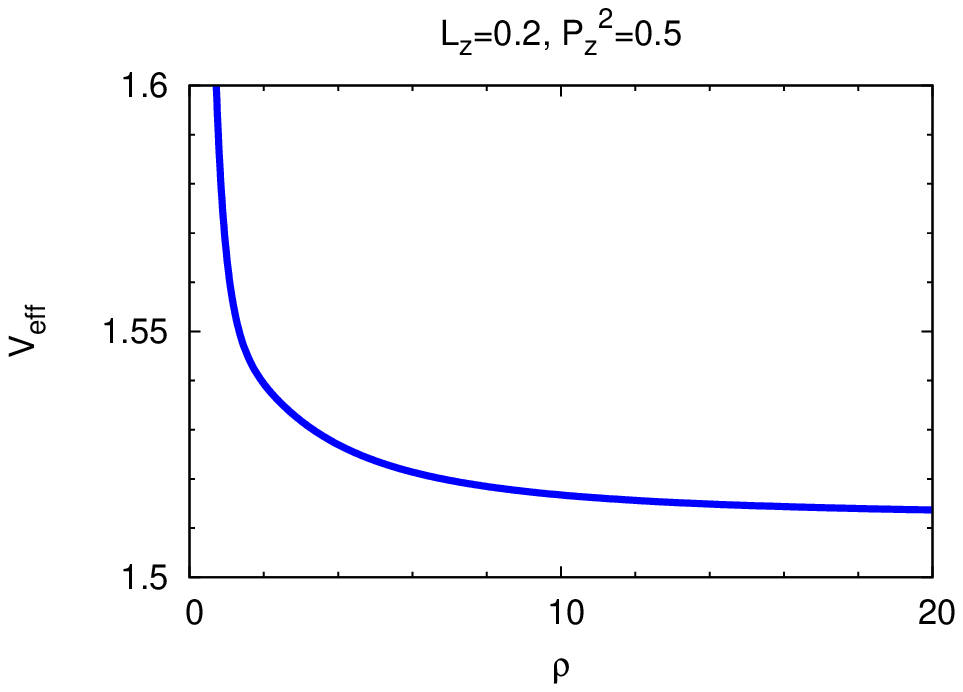}}
\subfigure[][$L_z^2=0.0, P_z^2=125$]{\label{c3_pot5}\includegraphics[width=5.0cm]{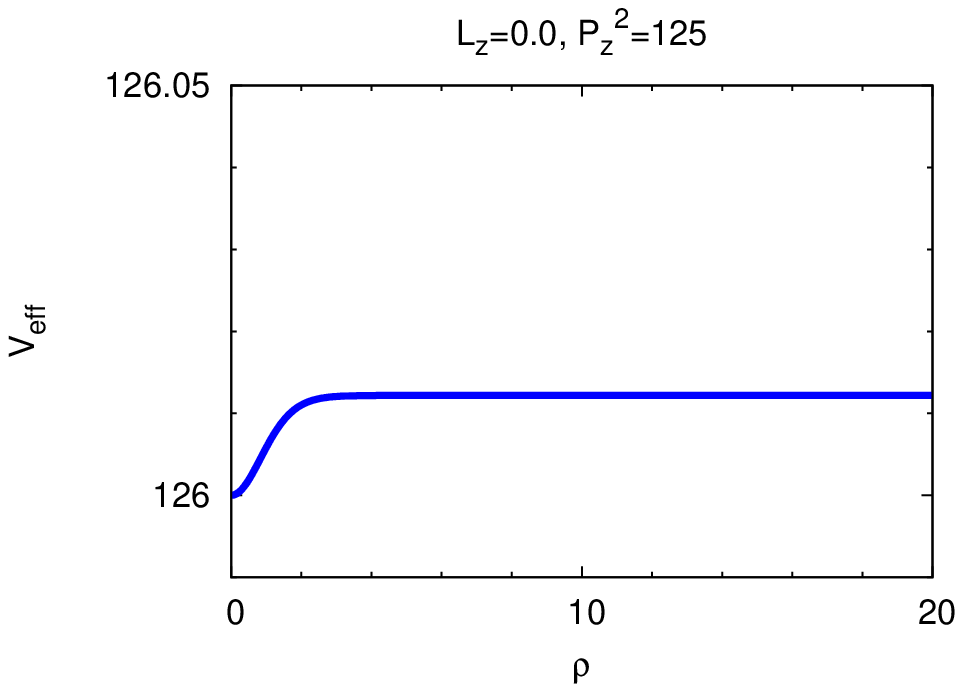}}
\subfigure[][$L_z^4=5.0 \cdot 10^{-5}, P_z^2=125$]{\label{c3_pot6}\includegraphics[width=5.0cm]{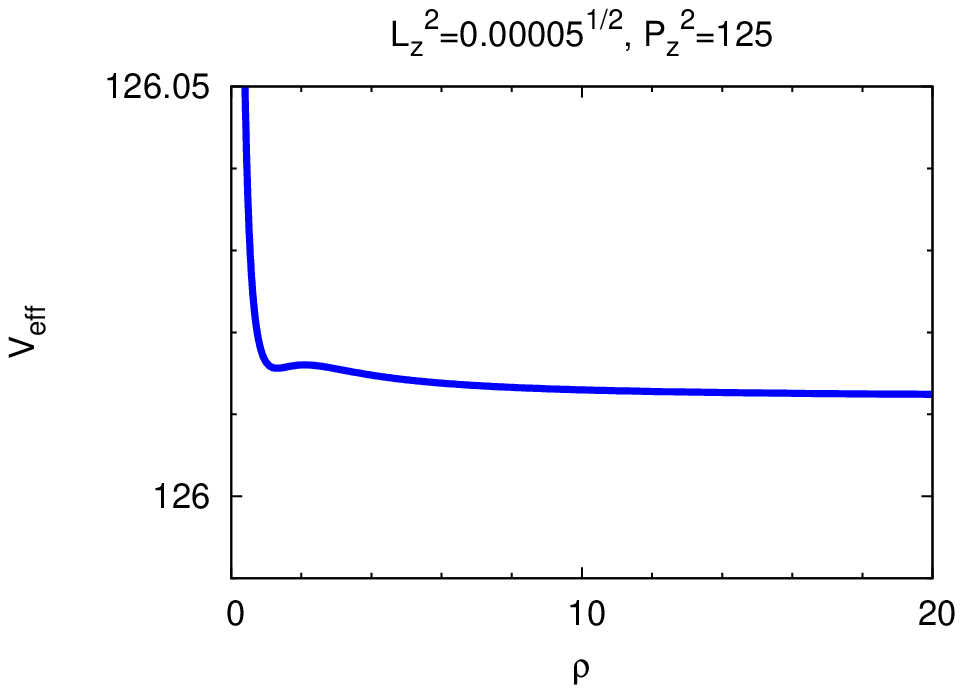}}
\subfigure[][$L_z^2=5.0 \cdot 10^{-5}, P_z^2=125$]{\label{c3_pot7}\includegraphics[width=5.0cm]{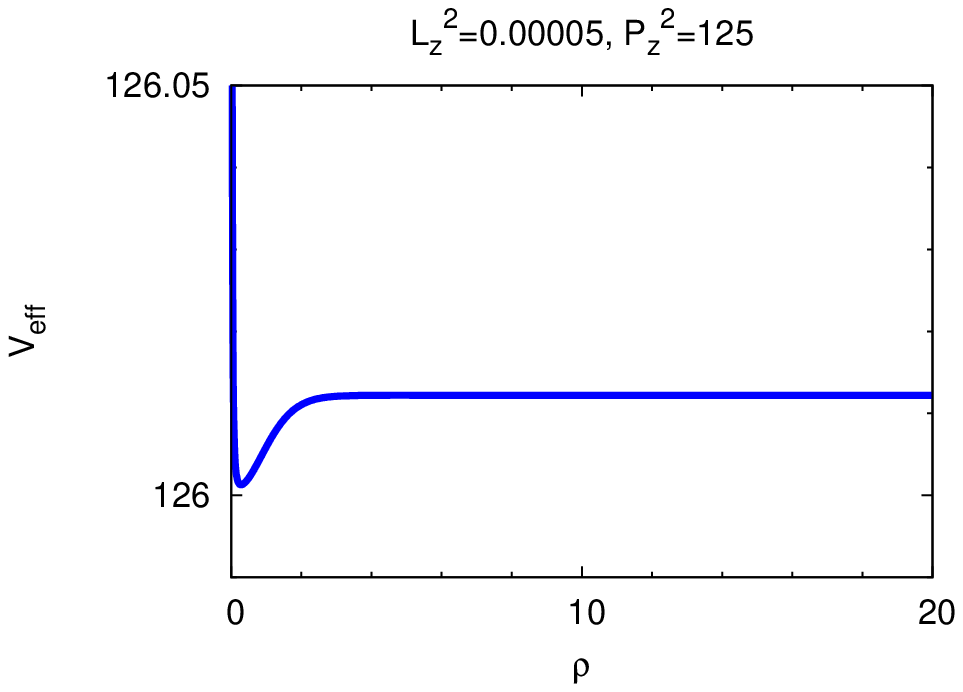}}
\subfigure[][$L_z=0.05, P_z^2=125$]{\label{c3_pot8}\includegraphics[width=5.0cm]{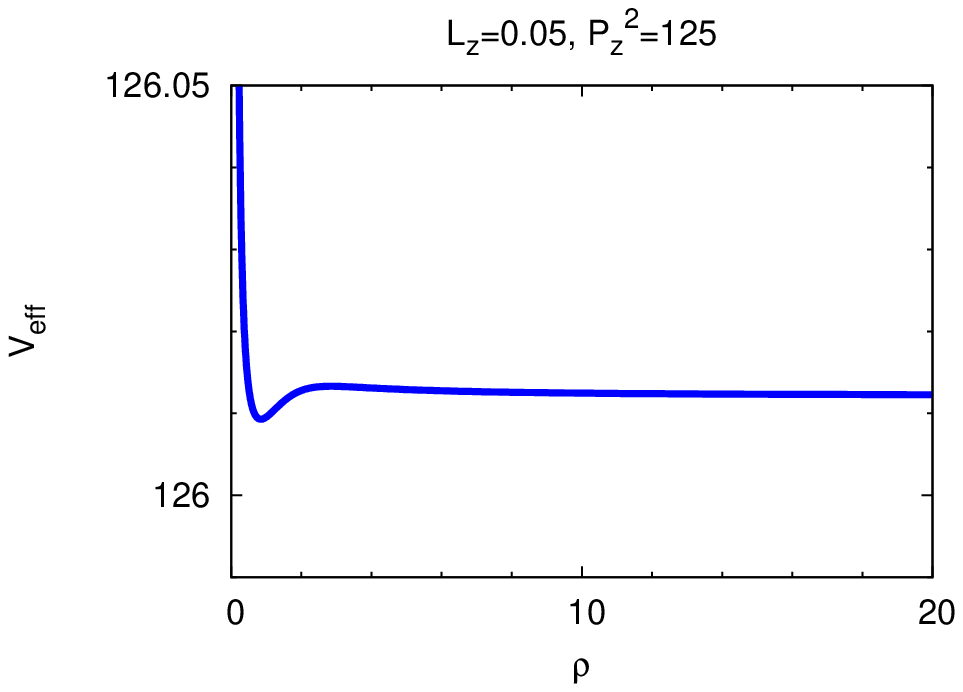}}
\subfigure[][$L_z=0.2, P_z^2=125$]{\label{c3_pot9}\includegraphics[width=5.0cm]{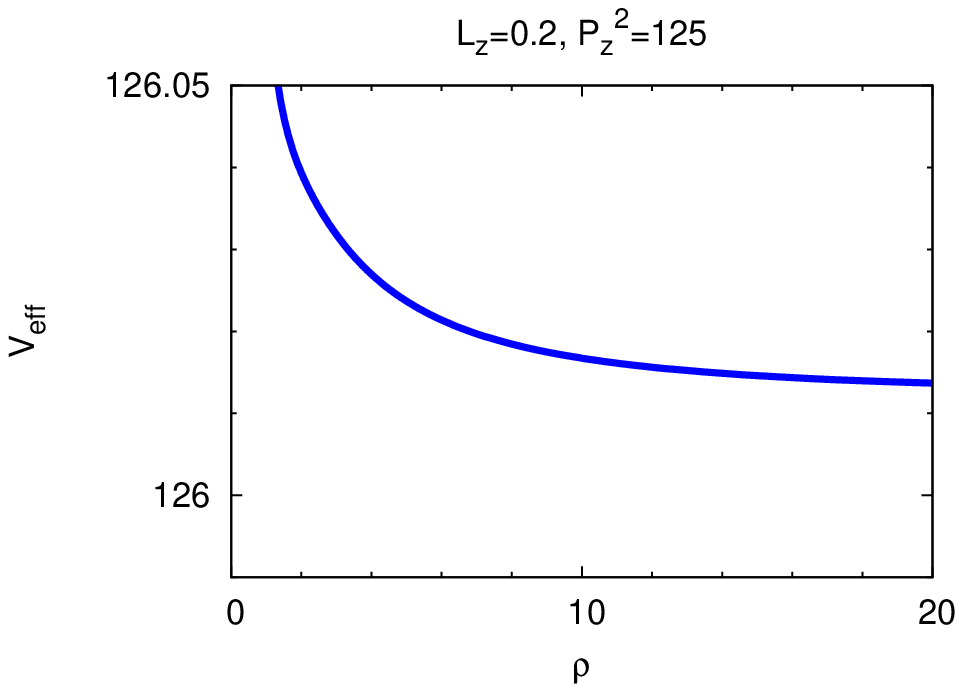}}
\end{center}
\caption{We show the effective potential for $n=m=1, \beta_1=\beta_2=1.8, \alpha=0.001, g=1.0, q=1.0$ \label{c3pots}.}
\end{figure*}

For $\beta_1 < 2$, $\beta_2 < 2$ bound orbits exist already for $\alpha=0$ \cite{hartmann_sirimachan}. 
For $\alpha\neq 0$ and $\beta_1 < 2$, $\beta_2 <2$ the effective potential is shown in Fig.\ref{c4pot_Lz2=0.0005}. In the following we are interested
in seeing the influence of the angular momentum $L_z$ and the linear momentum $P_z$ on the form of the
effective potential and consequently on the orbits.  In Fig.\ref{c3pots} we show the effective potential
for $\beta_1=\beta_2=1.8$, $\alpha=0.001$, $n=m=1$ and different values of $L_z$ and $P_z$. 
Again we observe that increasing the angular momentum from zero, the potential starts to form local
extrema such that bound orbits can exist. This is shown for $P_z^2=0.5$ in 
Fig.s \ref{c3_pot1}-\ref{c3_pot3} and
for $P_z^2=125$ in Fig.s \ref{c3_pot5}-\ref{c3_pot8}. For angular
momenta too large no bound orbits exist since the extrema have disappeared (see Fig. \ref{c3_pot4} for $P_z^2=0.5$ and
Fig.\ref{c3_pot9} for $P_z^2=125$.). This is connected to the fact that the repulsive centrifugal
force acting on these particles is too large to be balanced by the attractive gravitational force.
The extrema of the potential~\eqref{effective_pot} depend only on $L_z$. 
For constant $L_z$ and growing $P_z$ the qualitative structure of the plot
 is not changing: the types of orbits are preserved and the graph is moving upwards.

\begin{figure*}
\subfigure[$L_z^2=0.0, P_z^2=4.0$]{\label{c7_pot1}
\includegraphics[width=7cm]{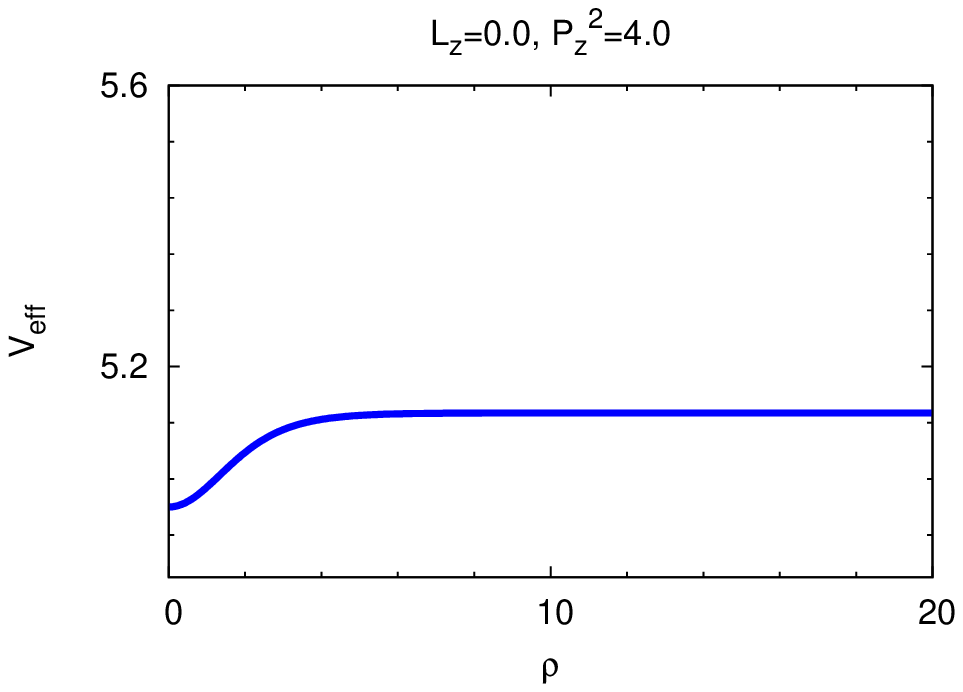}}
\subfigure[$L_z^2=5.0 \cdot 10^{-4}, P_z^2=4.0$]{\label{c7_pot2}
\includegraphics[width=7cm]{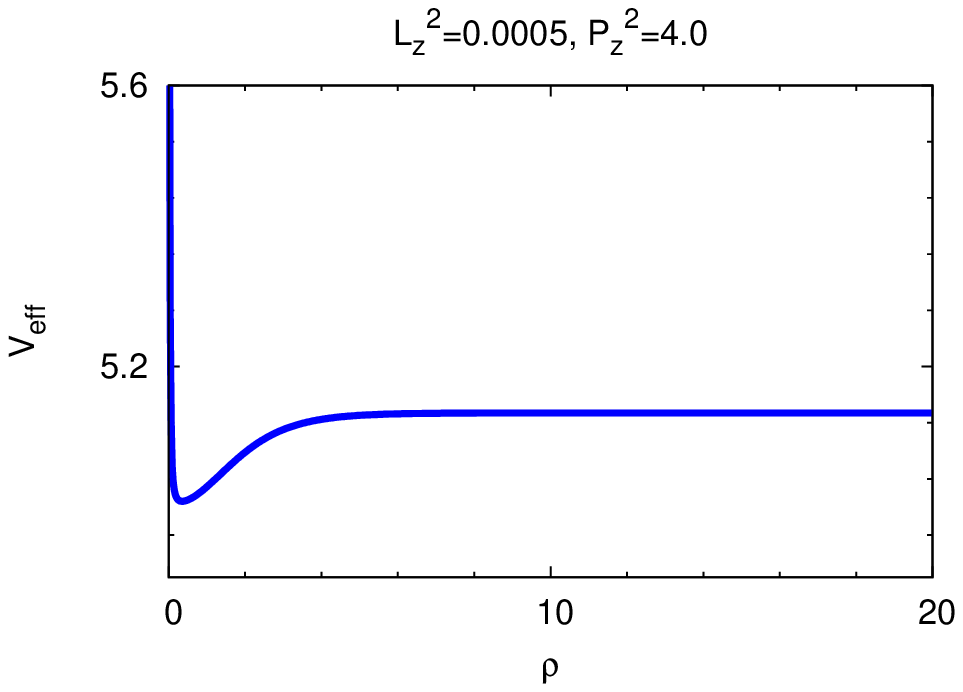}}\\
\subfigure[$L_z^2=5.0 \cdot 10^{-2}, P_z^2=4.0$]{\label{c7_pot3}
\includegraphics[width=7cm]{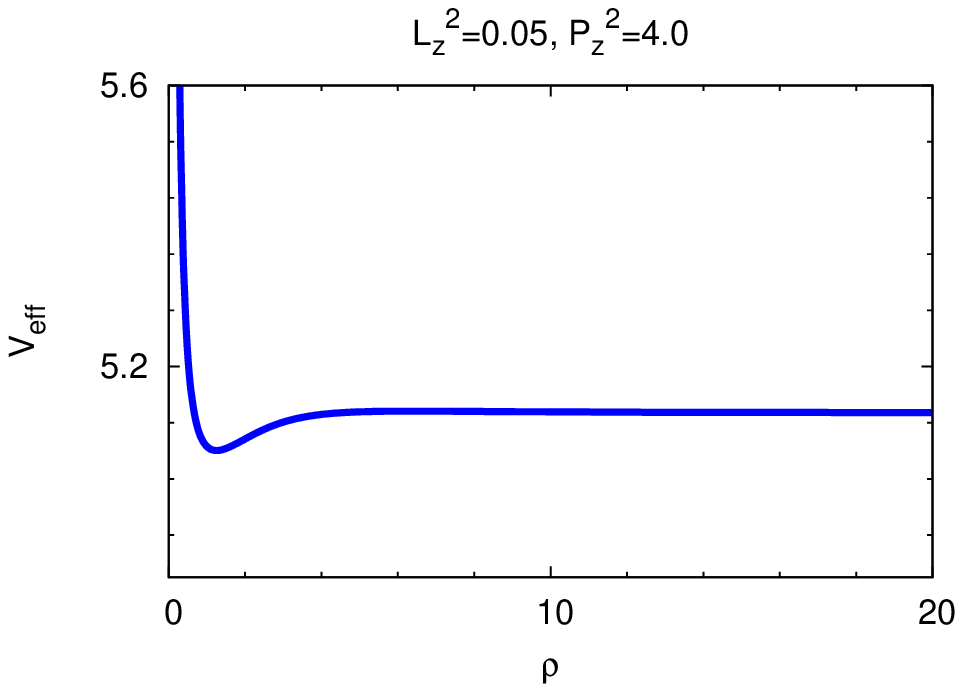}}
\subfigure[$L_z=0.5, P_z^2=4.0$]{\label{c7_pot4}\includegraphics[width=7cm]{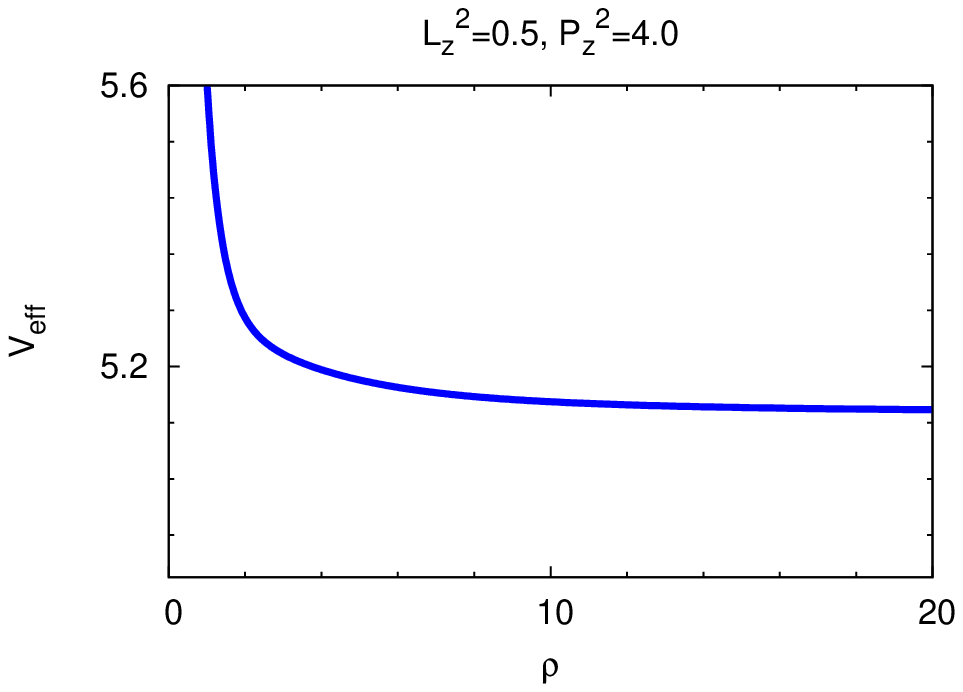}}
\caption{We show the effective potential for $n=1, m=-1, \beta_1=\beta_2=0.2, \alpha=0.001, g=1.0, q=1.0$ 
and different values of $L_z$ and $P_z$. \label{c7pots}}
\end{figure*}

Next we have investigated how the windings influence our results. Here, 
we want to put the emphasize on two strings with opposite windings 
and hence oppositely orientated magnetic fields interacting with each other.
Our results for the effective potential in the case $n=1$, $m=-1$, $\beta_1=\beta_2=0.2$ 
and $\alpha=0.001$ are shown in Fig.\ref{c7pots}. 
To compare, we also plotted the potential for $n=1$, $m=1$, $\beta_1=\beta_2=0.2$ 
and $\alpha=0.001$ and the same values of $L_z$ and $P_z$ in Fig.\ref{c4pot_Lz2=0.0005} 
(cf. Fig.\ref{c7_pot2}). There seems to be no qualitative 
difference between the plots for negative and positive windings, which
can also be noted when plotting the corresponding orbits (see next Section).
However, we expect that the winding would effect 
the orbits of charged test particles. This assumption 
comes from the analogy with the motion of charged particles 
in the field in charged Reissner-Nordstr\"om black-hole~\cite{GK2011}, 
where the cross-interaction between the electric and 
magnetic charges of the test particles and the gravitating source 
tear the test particle from the equatorial plane and make it move on a cone. This
is currently under investigation.

To conclude we observe that cosmic strings interacting via their magnetic fields
can capture test particles on
bound orbits if the energy and angular momentum of the test particle is not too large. 
We also observe that increasing the value of $\alpha$ we have to increase the value of 
$L_z^2$ by the same order of magnitude to find bound orbits. E.g. for $\alpha=0.01$ 
we find that bound orbits exist for $L_z^2$ on the order of $10^{-4}$.

\subsection{Examples of orbits}

\subsubsection{Massless test particles}

\begin{figure*}[h]
\begin{center}
\subfigure[][escape orbit]{\label{null1}\includegraphics[width=7.0cm]{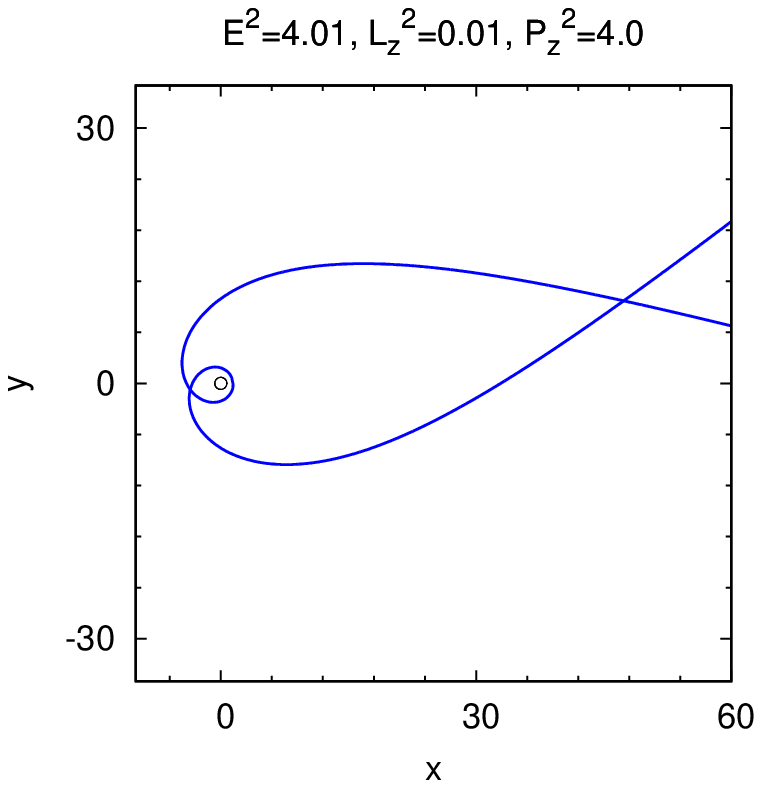}}
\subfigure[][3-d escape orbit]{\label{null2}\includegraphics[width=9.0cm]{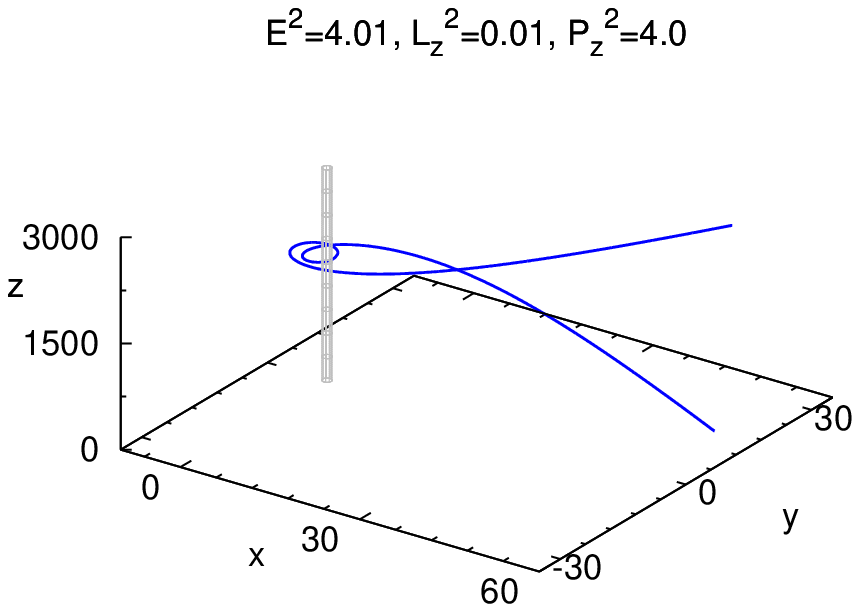}} \\
\subfigure[][escape orbit]{\label{null3}\includegraphics[width=7.0cm]{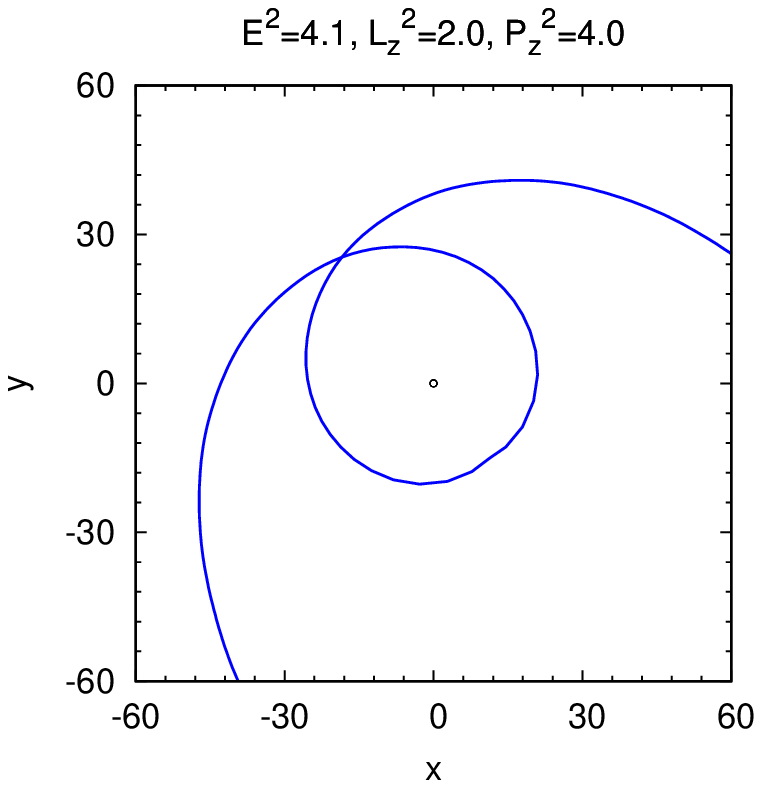}}
\subfigure[][3-d escape orbit]{\label{null4}\includegraphics[width=9.0cm]{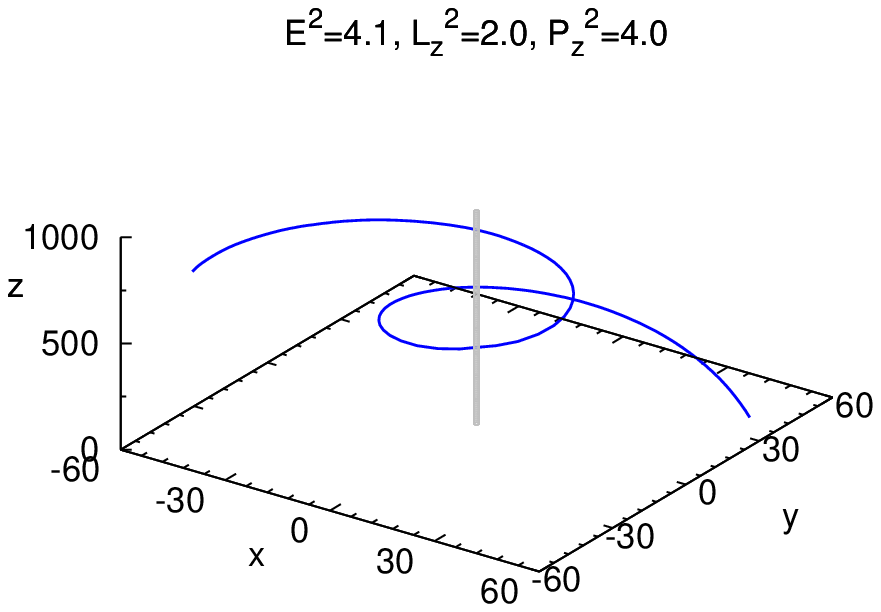}}
\end{center}
\caption{We show the null geodesics for $n=m=1, \beta_1=\beta_2=2.0, \alpha=0.001, g=1.0, q=1.0$ \label{photon_orb1}.}
\end{figure*}

Example of orbits for massless test particles is presented in Fig.\ref{photon_orb1}. As expected there are no bound orbits.
We observe that the test particle encircles the string before moving off again to infinity. This is not possible
in the case of infinitely thin cosmic strings and is due to the finite width of the core of the string.
For $E^2=4.01$ and $L_z^2=0.01$ in the Fig.s\ref{null1},\subref{null2} the turning point of the motion is closer to the string axis as for
$E^2=4.1$ and $L_z^2=2$ in the Fig.s\ref{null3},\subref{null4}. Hence the particle can interact stronger with the cosmic string via the curvature
of space-time.

\subsubsection{Massive test particles}
\begin{figure*}[t]
\begin{center}
\subfigure[bound orbit]{\label{c1_pot2_orbE2_5.000134b}
\includegraphics[width=7cm]{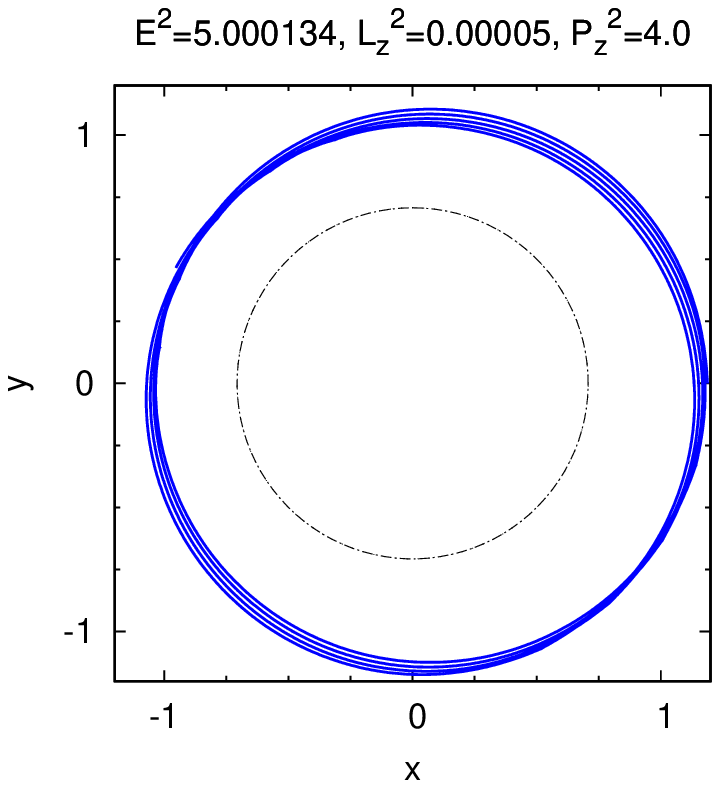}}
\subfigure[3-d bound orbit]{\label{c1_pot2_orbE2_5.000134b_3d}
\includegraphics[width=9cm]{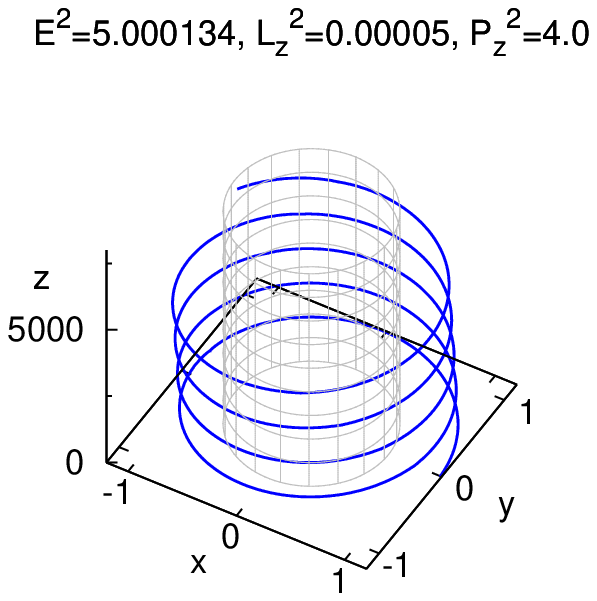}} \\
\subfigure[escape orbit]{\label{c1_pot2_orbE2_5.000134e}
\includegraphics[width=7cm]{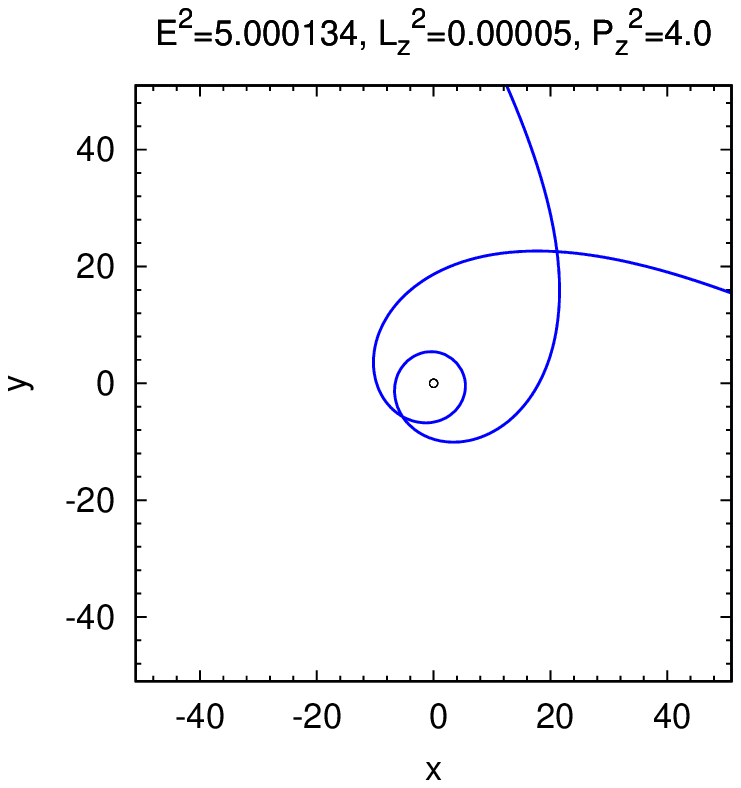}}
\subfigure[3-d escape orbit]{\label{c1_pot2_orbE2_5.000134e_3d}
\includegraphics[width=9cm]{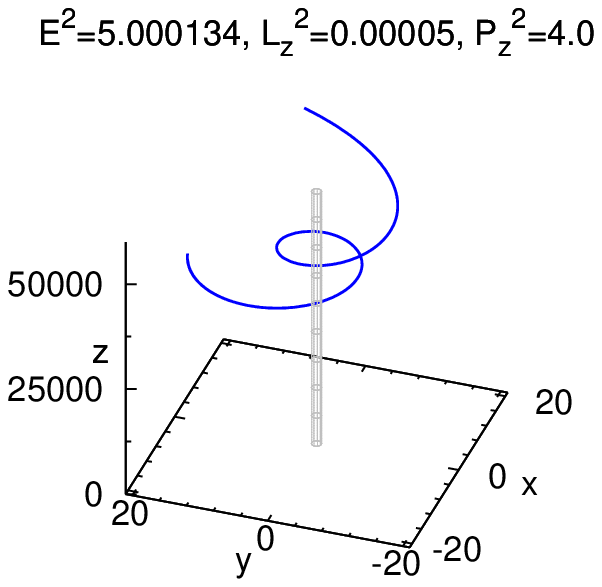}} 
\end{center}
\caption{We show a bound orbit in the $x$-$y$-plane (a) and in the 3-d (b), respectively, as well as an escape orbit
in the $x$-$y$-plane (c) and in 3-d (d), respectively of a massive test particle ($\varepsilon=1$) with 
$E^2=5.000134$, $L_z^2=5\cdot 10^{-5}$ and $P_z^2=4$ in the space-time of two Abelian-Higgs strings
interacting via their magnetic fields with $n=1, m=1, \beta_1=2.0, \beta_2=2.0, \alpha=0.001, g=1.0, q=1.0$ 
\label{c1pots_orbits}. The dashed circle and grey cylinder, respectively indicate the string core.}
\end{figure*}

\begin{figure*}[t]
\begin{center}
\subfigure[][bound orbit]
{\label{c1_pot2_orbE2_5.000146b}\includegraphics[width=7cm]{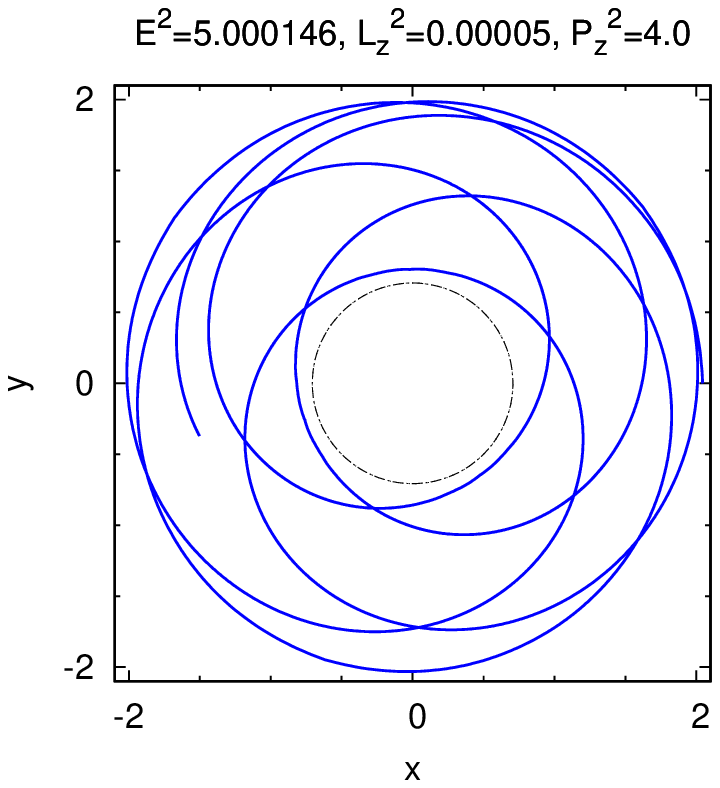}}
\subfigure[][3-d bound orbit ]
{\label{c1_pot2_orbE2_5.000146b_3d}\includegraphics[width=9cm]{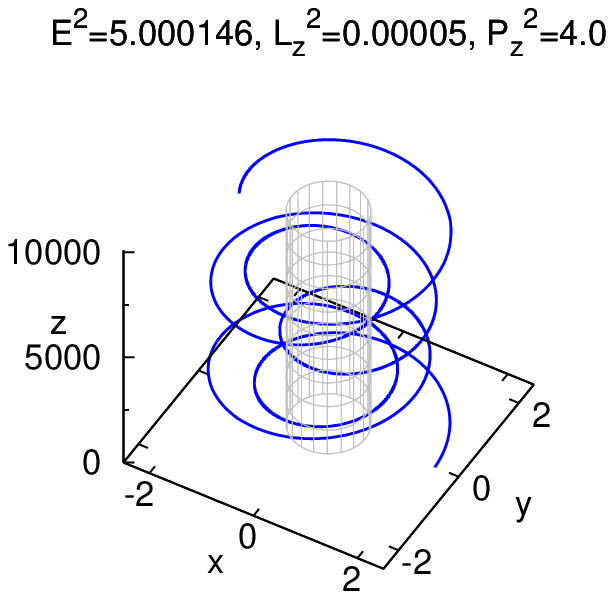}} \\
\subfigure[][escape orbit]
{\label{c1_pot2_orbE2_5.000146e}\includegraphics[width=7cm]{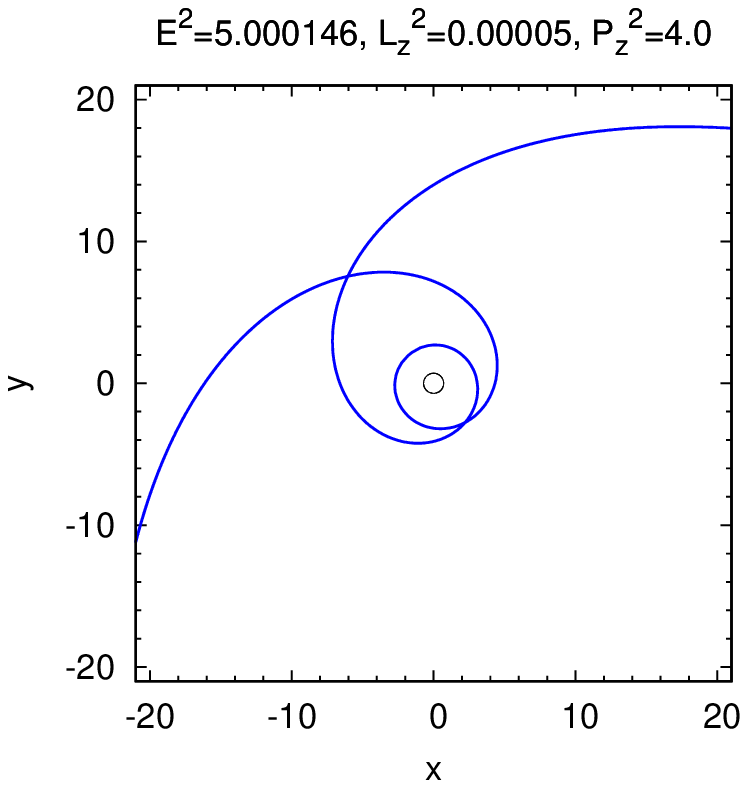}}
\subfigure[][3-d escape orbit]
{\label{c1_pot2_orbE2_5.000146e_3d}\includegraphics[width=9cm]{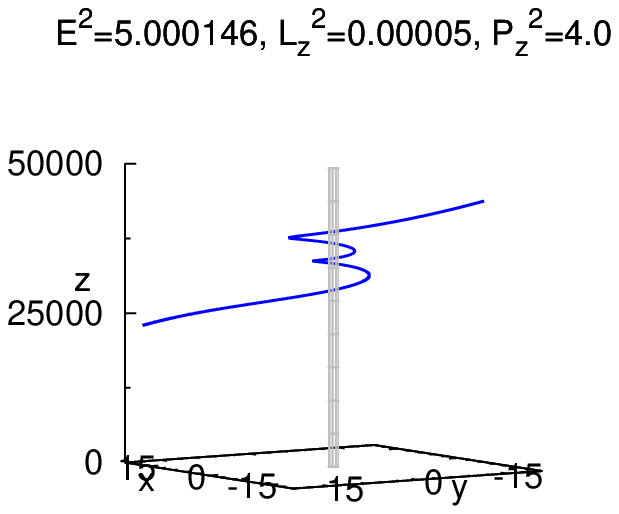}} 
\end{center}
\caption{
We show a bound orbit in the $x$-$y$-plane (a) and in the 3-d (b), respectively, as well as an escape orbit
in the $x$-$y$-plane (c) and in 3-d (d), respectively of a massive test particle ($\varepsilon=1$) with 
$E^2=5.000146$, $L_z^2=5\cdot 10^{-5}$ and $P_z^2=4$ in the space-time of two Abelian-Higgs strings
interacting via their magnetic fields with $n=1, m=1, \beta_1=2.0, \beta_2=2.0, \alpha=0.001, g=1.0, q=1.0$ 
\label{c1pots_orbits2}. The dashed circle and grey cylinder, respectively indicate the string core.}
\end{figure*}

\begin{figure*}[t]
\begin{center}
\subfigure[][escape orbit]{\label{c1_pot2_orbE2_5.000147e}
\includegraphics[width=7cm]{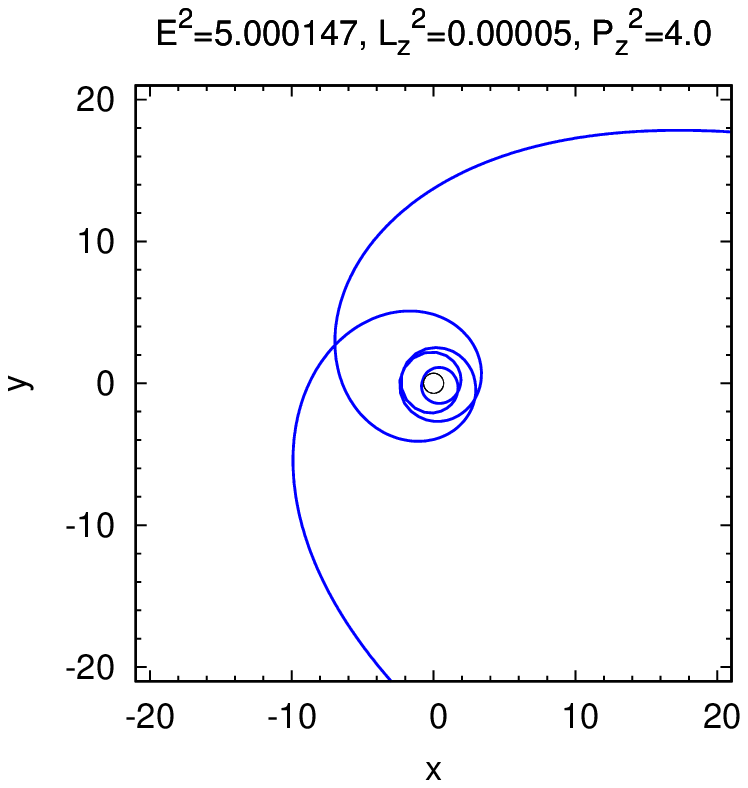}}
\subfigure[][3-d escape orbit]{\label{c1_pot2_orbE2_5.000147e_3d}
\includegraphics[width=9cm]{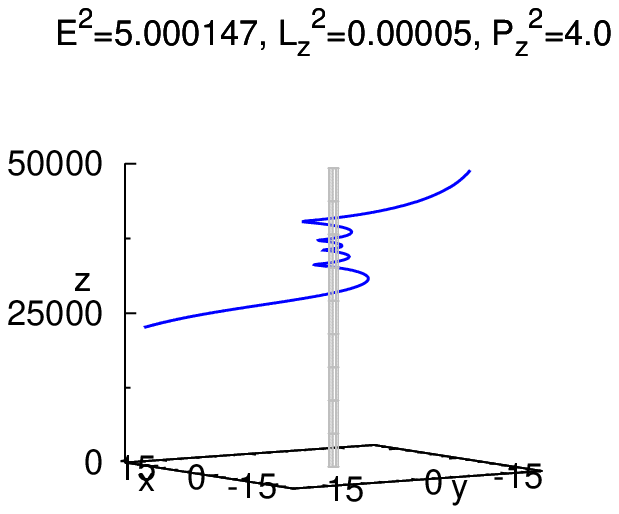}}
\end{center}
\caption{We show an escape orbit
in the $x$-$y$-plane (a) and in 3-d (b), respectively of a massive test particle ($\varepsilon=1$) with 
$E^2=5.000147$, $L_z^2=5\cdot 10^{-5}$ and $P_z^2=4$ in the space-time of two Abelian-Higgs strings
interacting via their magnetic fields with $n=1, m=1, \beta_1=2.0, \beta_2=2.0, \alpha=0.001, g=1.0, q=1.0$ 
\label{c1pots_orbits3}. The dashed circle and grey cylinder, respectively indicate the string core.}
\end{figure*}

We show the orbits corresponding to the potentials in Fig.\ref{c1pots} 
in Fig.s \ref{c1pots_orbits}-\ref{c1pots_orbits3} for different values of ${\cal E}=E^2$. 
Let us first consider the case $L_z^2=5\cdot 10^{-5}$ (see Fig.\ref{c1_pot2}). 
For $E^2=5.000134$ we have a bound and an escape orbit. These are shown in Fig.s \ref{c1_pot2_orbE2_5.000146b}-
\ref{c1_pot2_orbE2_5.000146b_3d} and Fig.s \ref{c1_pot2_orbE2_5.000146e}-\ref{c1_pot2_orbE2_5.000146e_3d}, respectively.
We give the orbit projected onto the $x$-$y$-plane (Fig.\ref{c1_pot2_orbE2_5.000146b} and Fig.\ref{c1_pot2_orbE2_5.000146e})
as well as the orbit in 3 dimensions (Fig.\ref{c1_pot2_orbE2_5.000146b_3d} and Fig.\ref{c1_pot2_orbE2_5.000146e_3d}).
The bound orbit is nearly circular in the  $x$-$y$-plane. This is related
to the fact that the energy value is close to the minimum of the potential. This bound orbit is hence close
to a stable circular orbit. On the escape orbit, the particles encircles the string before moving off again
to infinity. 

Increasing the energy, we find that the bound orbit moves away from circularity stronger and stronger and starts
to develop a large perihelion shift. This is seen in Fig.s \ref{c1_pot2_orbE2_5.000146b} - 
\ref{c1_pot2_orbE2_5.000146b_3d}
where we show a bound for $E^2=5.000146$. For the same value of $E^2$ an escape orbit exists.
This is given in Fig.s \ref{c1_pot2_orbE2_5.000146e}-\ref{c1_pot2_orbE2_5.000146e_3d}. Qualitatively, the escape
orbit looks similar to the one for smaller energy, however, we observe that the test particle comes
closer to the string core.

Increasing the value of $E^2$ beyond the value of the maximum of the effective potential we find that only
escape orbits exist and no bound orbits exist. This is seen in Fig.\ref{c1pots_orbits3} where we show the escape orbit for 
$E^2=5.000147$. We observe that an increase in energy leads to a stronger interaction of the particle
with the string in the sense that it encircles the string more often before moving away to infinity again.

\begin{figure*}[t]
\begin{center}
\subfigure[][bound orbit]{\label{orb1_b}\includegraphics[width=7cm]
{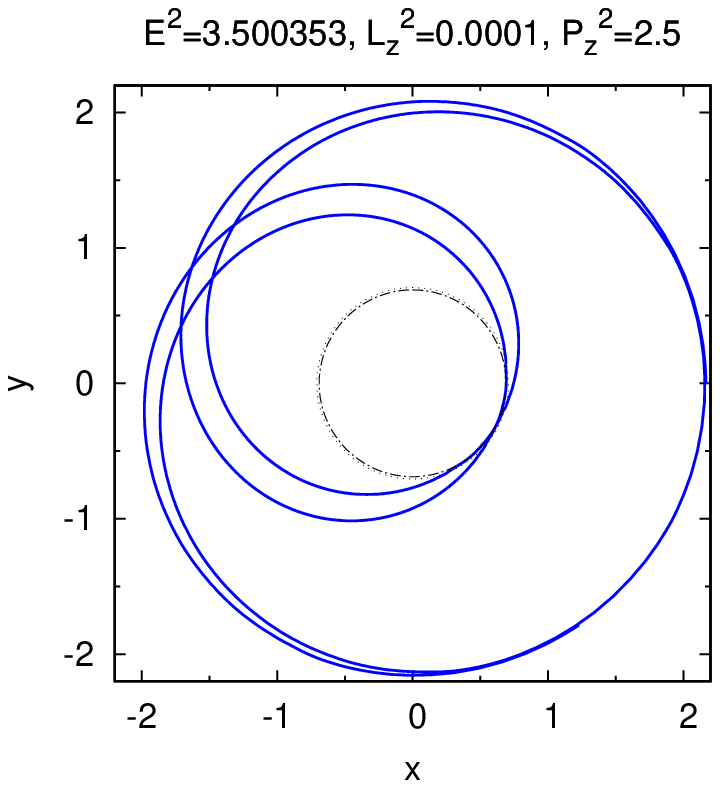}}
\subfigure[][3-d bound orbit]{\label{orb1_b3d}
\includegraphics[width=9cm]{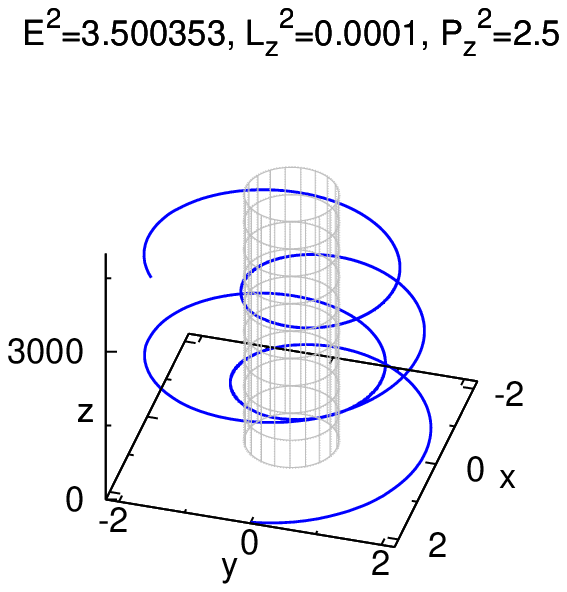}} \\
\subfigure[][escape orbit]{\label{orb1_e}
\includegraphics[width=7cm]{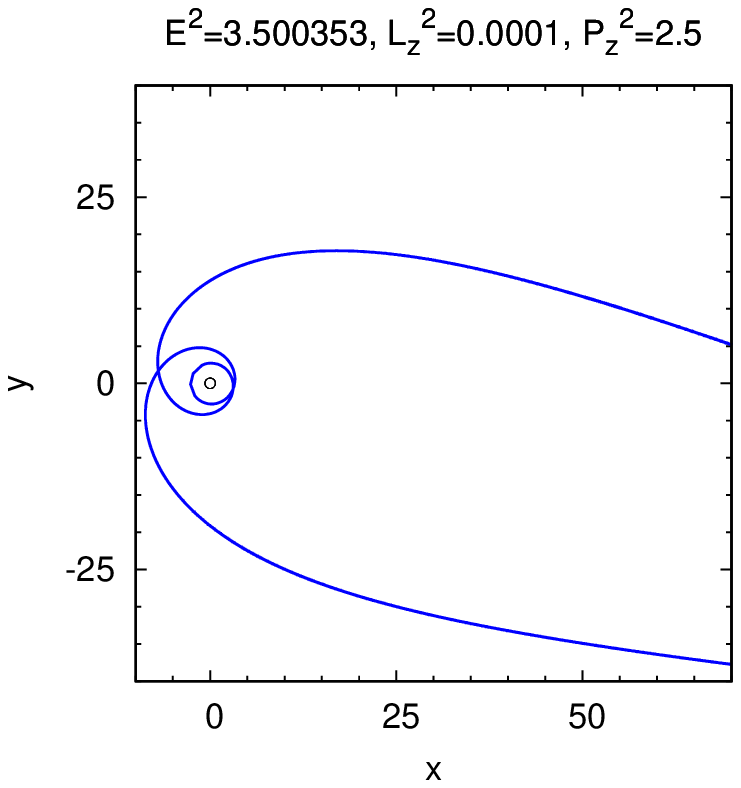}} 
\subfigure[][3-d escape orbit]{\label{orb1_e3d}
\includegraphics[width=9cm]{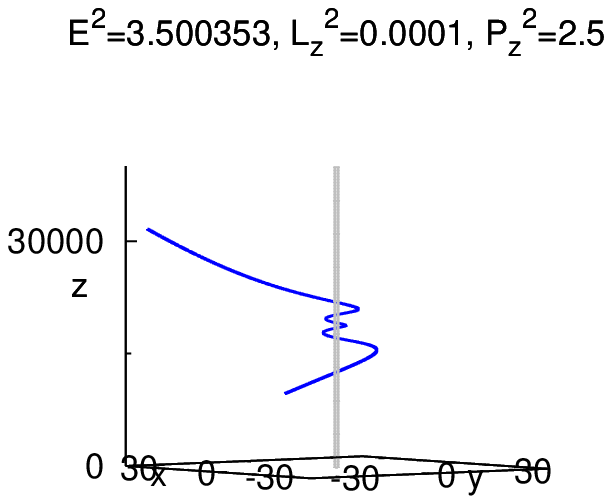}}
\end{center}
\caption{
We show a bound orbit in the $x$-$y$-plane (a) and in the 3-d (b), respectively, as well as an escape orbit
in the $x$-$y$-plane (c) and in 3-d (d), respectively of a massive test particle ($\varepsilon=1$) with 
$E^2=3.500353$, $L_z^2=10^{-4}$ and $P_z^2=2.5$ in the space-time of two Abelian-Higgs strings
interacting via their magnetic fields with $n=1, m=1, \beta_1=\beta_2=2.1, \alpha=0.05, g=1.0, q=1.0$. 
The dashed circle and grey cylinder, respectively indicate the string core.\label{orbits_b1b2=2.1}}
\end{figure*}

\begin{figure*}[t]
\begin{center}
\subfigure[][escape orbit]{\label{orb2_e}
\includegraphics[width=4cm]{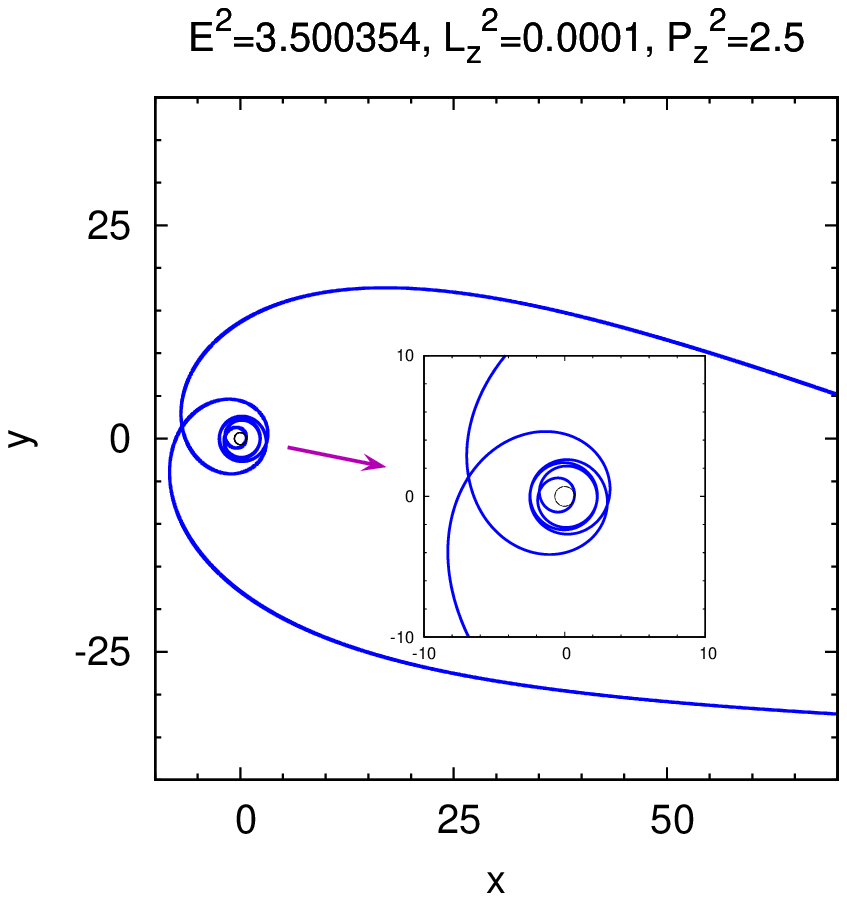}} 
\subfigure[][3-d escape orbit]{\label{orb2_e3d}
\includegraphics[width=9cm]{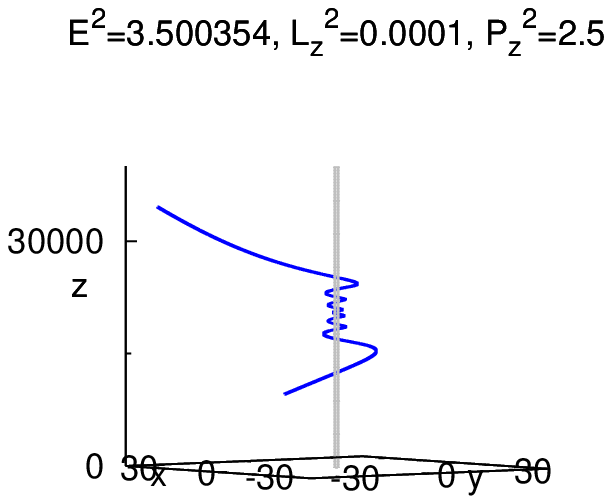}}
\end{center}
\caption{
We show an escape orbit
in the $x$-$y$-plane (a) and in 3-d (b), respectively of a massive test particle ($\varepsilon=1$) with 
$E^2=3.500354$, $L_z^2=10^{-4}$ and $P_z^2=2.5$ in the space-time of two Abelian-Higgs strings
interacting via their magnetic fields with $n=m=1, \beta_1=\beta_2=2.1, \alpha=0.05, g=1.0, q=1.0$. 
The dashed circle and grey cylinder, respectively indicate the string core.\label{orbits_b1b2=2.1_2}}
\end{figure*}

As already mentioned in the discussion on the effective potential bound orbits are not possible
for $\alpha=0$ and $\beta_1 \geq 2$, $\beta_2\geq 2$ \cite{hartmann_sirimachan}. In the example
just discussed, we have seen that bound orbits are possible for $\alpha > 0$ for $\beta_1=\beta_2=2$.
In the following, we will show examples of orbits for $\beta_1 =\beta_2 > 2$. 
For this, we choose $L_z^2 = 10^{-4}$, $P_z^2=2.5$, $\beta_1=\beta_2=2.1$ and $\alpha=0.05$. This
corresponds to the potential shown in Fig.\ref{b_1_2_21_pot}. For $E^2=3.500353$ we find
that a bound and an escape orbit exist. The bound orbit is shown in Fig.s \ref{orb1_b}-\ref{orb1_b3d}, while
the escape orbits is given in Fig.s \ref{orb1_e}-\ref{orb1_e3d}. On the bound orbit, the test particle
moves nearly on a circular orbit, however this orbit possesses an additional loop that touches
the string core. On the escape orbit, the test particle encircles again the string and moves back to infinity
on a path nearly parallel to the path that the particle originally came from.
From far this hence looks as if the test particle would be reflected by the string with angle $\approx \pi$.
Increasing the energy only escape orbits exist. This is shown for $E^2=3.500354$ in Fig.\ref{orbits_b1b2=2.1_2}. 
The test particle encircles the string more often before moving again to infinity as compared to the case
with smaller energy.

\begin{figure*}[t]
\begin{center}
\subfigure[][bound orbit]{\label{c3Lz_0.05_Pz2_125_E2_126.013b}\includegraphics[width=7cm]
{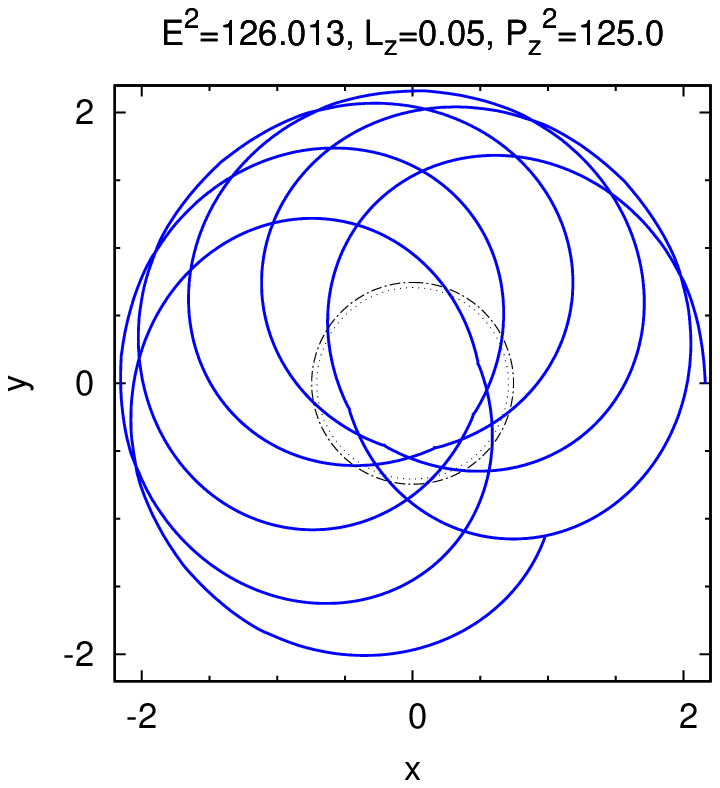}}
\subfigure[][3-d bound orbit]{\label{c3Lz_0.05_Pz2_125_E2_126.013b_3d}
\includegraphics[width=9cm]{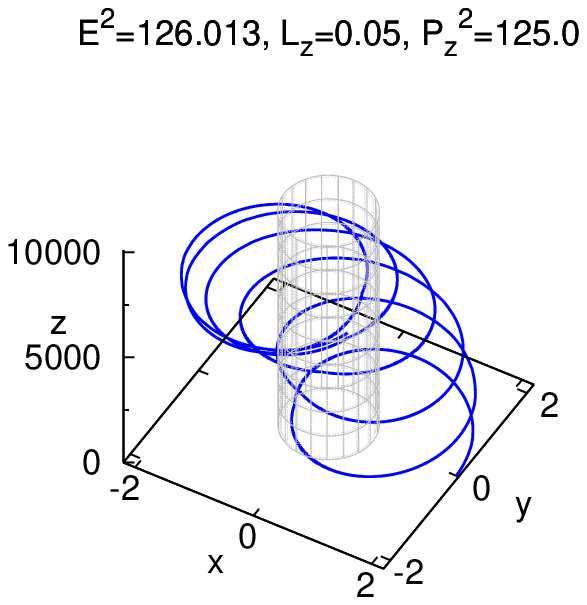}} \\
\subfigure[][escape orbit]{\label{c3Lz_0.05_Pz2_125_E2_126.013e}
\includegraphics[width=7cm]{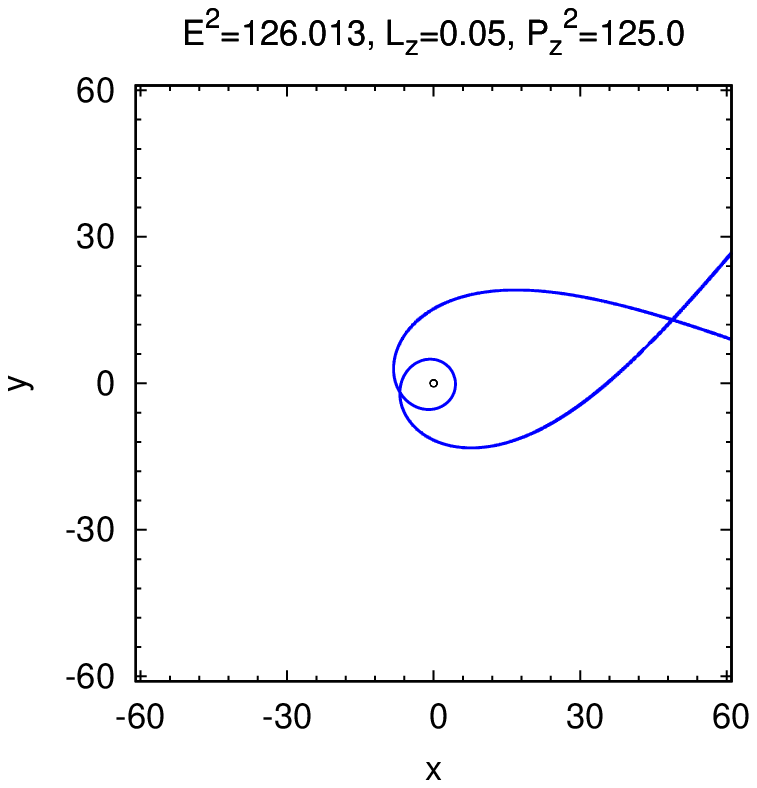}}
\subfigure[][3-d escape orbit]{\label{c3Lz_0.05_Pz2_125_E2_126.013e_3d}
\includegraphics[width=9cm]{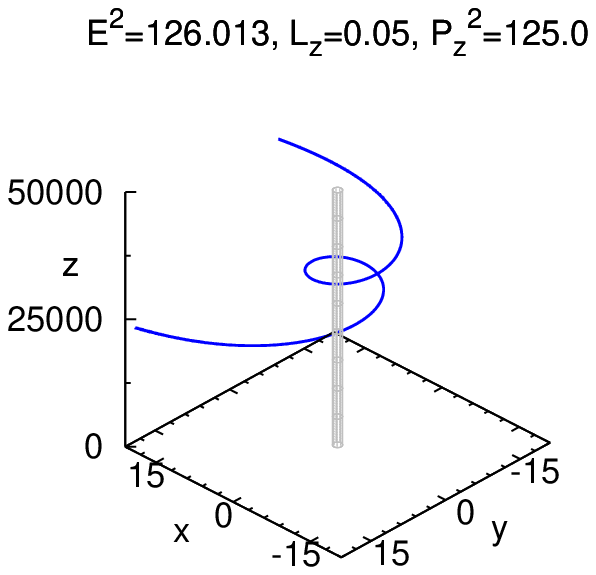}} 
\end{center}
\caption{
We show a bound orbit in the $x$-$y$-plane (a) and in the 3-d (b), respectively, as well as an escape orbit
in the $x$-$y$-plane (c) and in 3-d (d), respectively of a massive test particle ($\varepsilon=1$) with 
$E^2=5.000147$, $L_z=0.05$ and $P_z^2=125$ in the space-time of two Abelian-Higgs strings
interacting via their magnetic fields with $n=m=1, \beta_1=\beta_2=1.8, \alpha=0.001, g=1.0, q=1.0$. 
The dashed circle and grey cylinder, respectively indicate the string core\label{c3pots_orbits}.}
\end{figure*}

\begin{figure*}[t]
\begin{center}
\subfigure[][escape orbit]{\label{c3Lz_0.05_Pz2_125_E2_126.0133e}
\includegraphics[width=7cm]{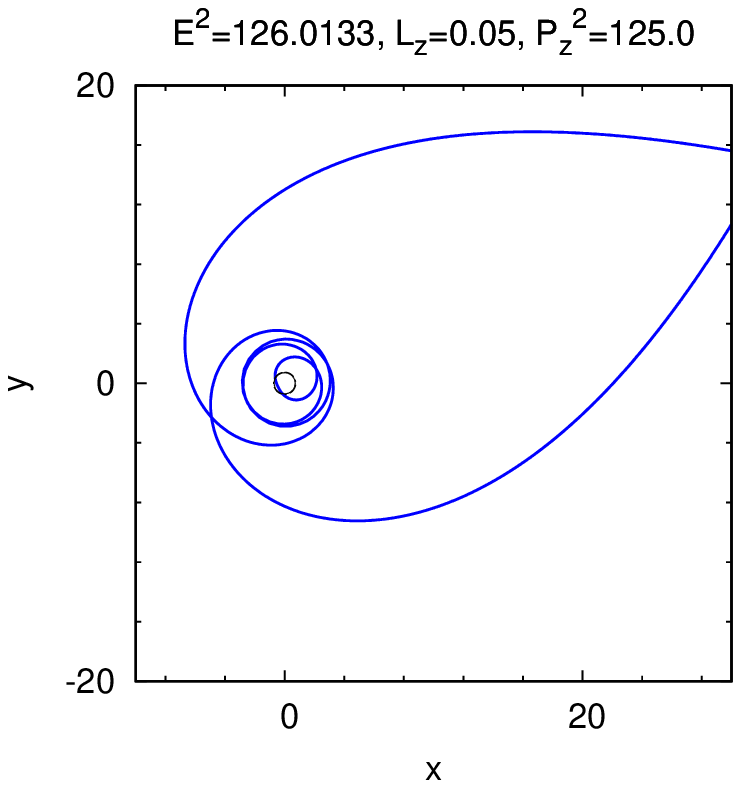}}
\subfigure[][3-d escape orbit]{\label{c3Lz_0.05_Pz2_125_E2_126.0133e_3d}
\includegraphics[width=9cm]{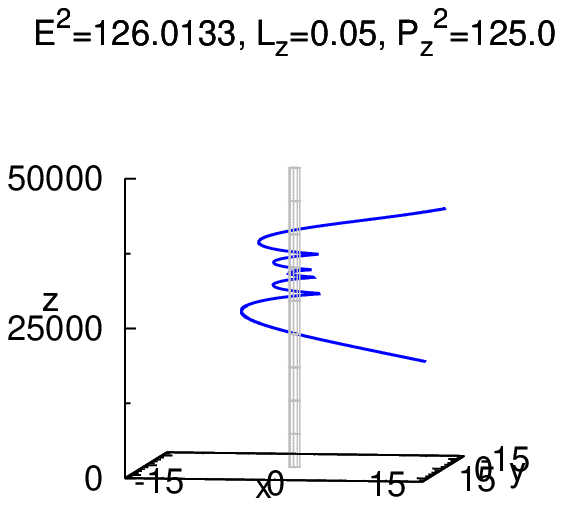}}
\end{center}
\caption{
We show an escape orbit
in the $x$-$y$-plane (a) and in 3-d (b), respectively of a massive test particle ($\varepsilon=1$) with 
$E^2=5.000147$, $L_z=0.05$ and $P_z^2=125$ in the space-time of two Abelian-Higgs strings
interacting via their magnetic fields with $n=m=1, \beta_1=\beta_2=1.8, \alpha=0.001, g=1.0, q=1.0$. 
The dashed circle and grey cylinder, respectively indicate the string core\label{c3pots_orbits2}.}
\end{figure*}

Finally, let us also discuss the case for $\beta_1 \leq 2$ and $\beta_2\leq 2$. In this
case, bound orbits are already possible for $\alpha=0$ \cite{hartmann_sirimachan}.
Here, we will choose much larger values for $E^2$, $L_z^2$ and $P_z^2$ in order
to show the influence of these parameters. 
We choose $\beta_1=\beta_2=1.8$, $n=m=1$, $\alpha=0.001$ and $P_z^2=125$, $L_z=0.05$.
The corresponding potential is shown in Fig.\ref{c3_pot8}. 
The orbits for $E^2=126.013$ are shown in Fig.\ref{c3pots_orbits}. In this case both a bound and an escape orbit
exist. The bound orbit is shown in Fig.s \ref{c3Lz_0.05_Pz2_125_E2_126.013b}-\ref{c3Lz_0.05_Pz2_125_E2_126.013b_3d}, while
the escape orbit is given in Fig.s \ref{c3Lz_0.05_Pz2_125_E2_126.013e}-\ref{c3Lz_0.05_Pz2_125_E2_126.013e_3d}.
The bound orbit possesses a perihelion shift and moves partially within the string core, i.e.
interacts directly with the region of space-time in which the matter fields have not yet
reached their vacuum values. Moreover on the escape orbit the trajectory forms two closed loops, one close to
the core of the string, one further out. 
Increasing the energy, only escape orbits exist. This is shown for 
$E^2=126.0133$ in Fig.s \ref{c3Lz_0.05_Pz2_125_E2_126.0133e}-\ref{c3Lz_0.05_Pz2_125_E2_126.0133e_3d}.
As for the cases discussed above, the increase in energy leads to a stronger interaction of the test particle
with the string and it encircles the string more often before moving again to infinity.



\begin{figure*}[t]
\begin{center}
\subfigure[][bound orbit]{\label{c7_pot2_orbE2_5.01b}
\includegraphics[width=7cm]{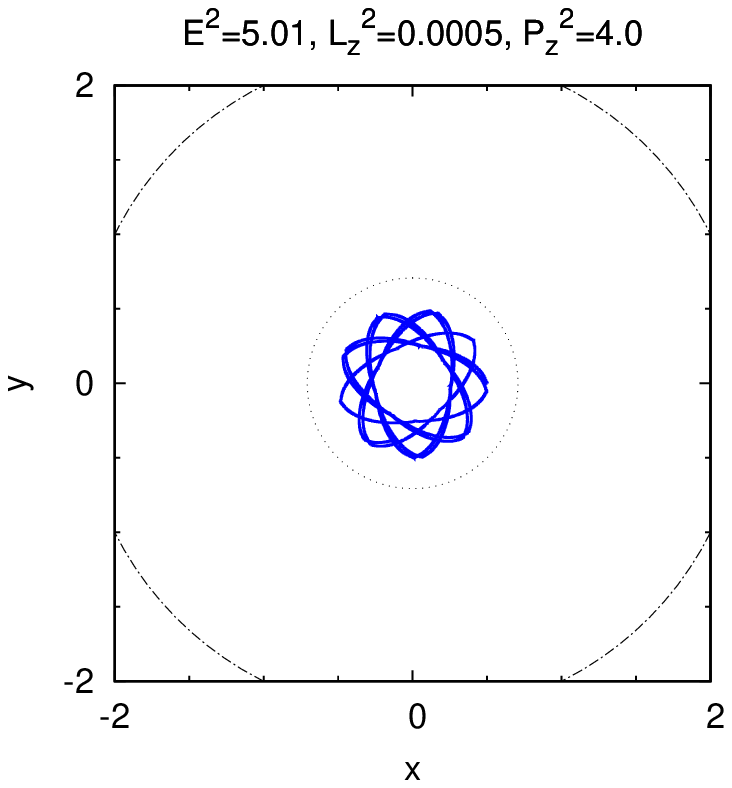}}
\subfigure[][3-d bound orbit]{\label{c7_pot2_orbE2_5.01b_3d}
\includegraphics[width=9cm]{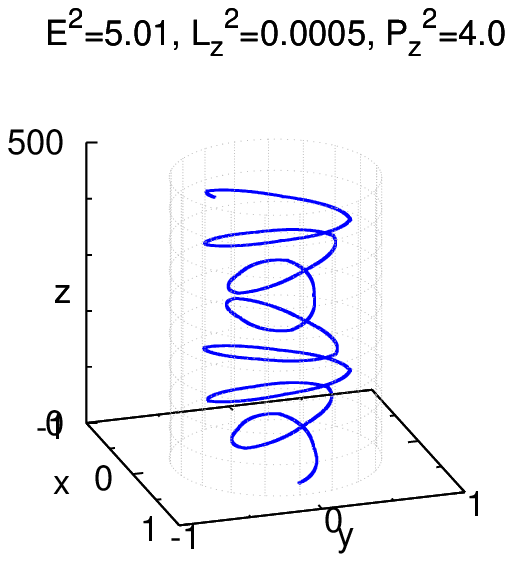}} 
\end{center}
\caption{
We show a bound orbit
in the $x$-$y$-plane (a) and in 3-d (b), respectively of a massive test particle ($\varepsilon=1$) with 
$E^2=5.01$, $L_z^2=0.0005$ and $P_z^2=4$ in the space-time of two Abelian-Higgs strings
interacting via their magnetic fields with $n=1, m=-1, \beta_1=\beta_2=0.2, \alpha=0.001, g=1.0, q=1.0$. 
The dashed circle and grey cylinder, respectively indicate the string core
\label{c7pots_orbits}.}
\end{figure*}

\begin{figure*}[t]
\begin{center}
\subfigure[][bound orbit]{\label{c7_pot2_orbE2_5.05b}
\includegraphics[width=7cm]{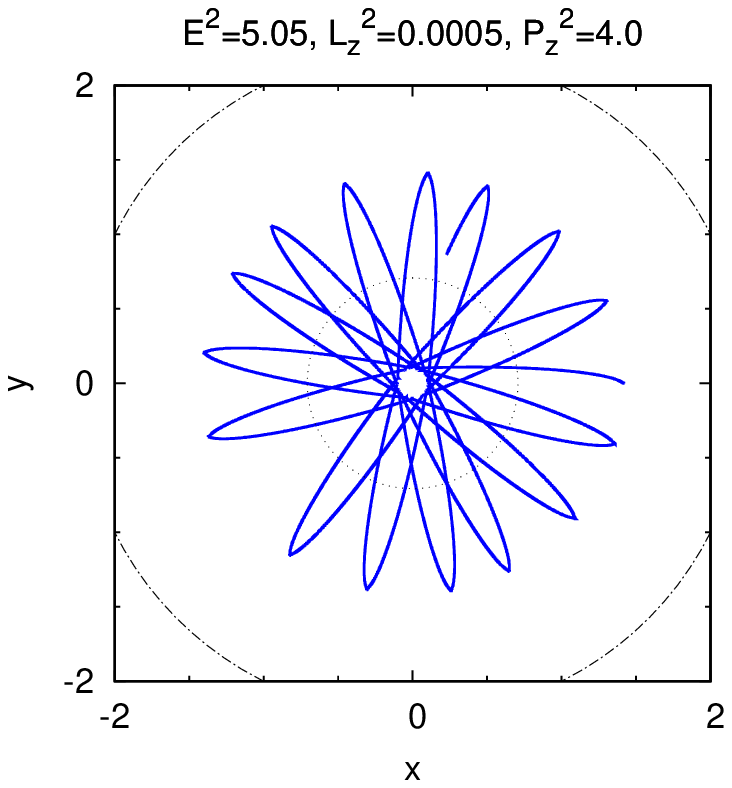}}
\subfigure[][3-d bound orbit]{\label{c7_pot2_orbE2_5.05b_3d}
\includegraphics[width=9cm]{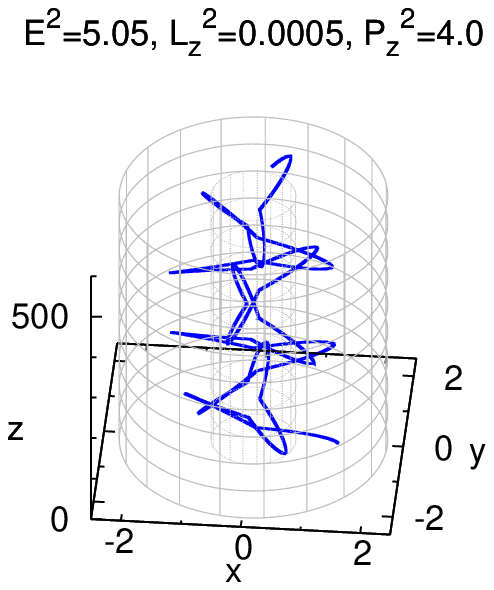}} 
\end{center}
\caption{
We show a bound orbit
in the $x$-$y$-plane (a) and in 3-d (b), respectively of a massive test particle ($\varepsilon=1$) with 
$E^2=5.05$, $L_z^2=0.0005$ and $P_z^2=4$ in the space-time of two Abelian-Higgs strings
interacting via their magnetic fields with $n=1, m=-1, \beta_1=\beta_2=0.2, \alpha=0.001, g=1.0, q=1.0$. 
The dashed circle and grey cylinder, respectively indicate the string core
\label{c7pots_orbits2}.}
\end{figure*}

\clearpage

\begin{figure*}[t]
\begin{center}
\subfigure[][bound orbit]{\label{c7_pot2_orbE2_5.13b}\includegraphics[width=7cm]{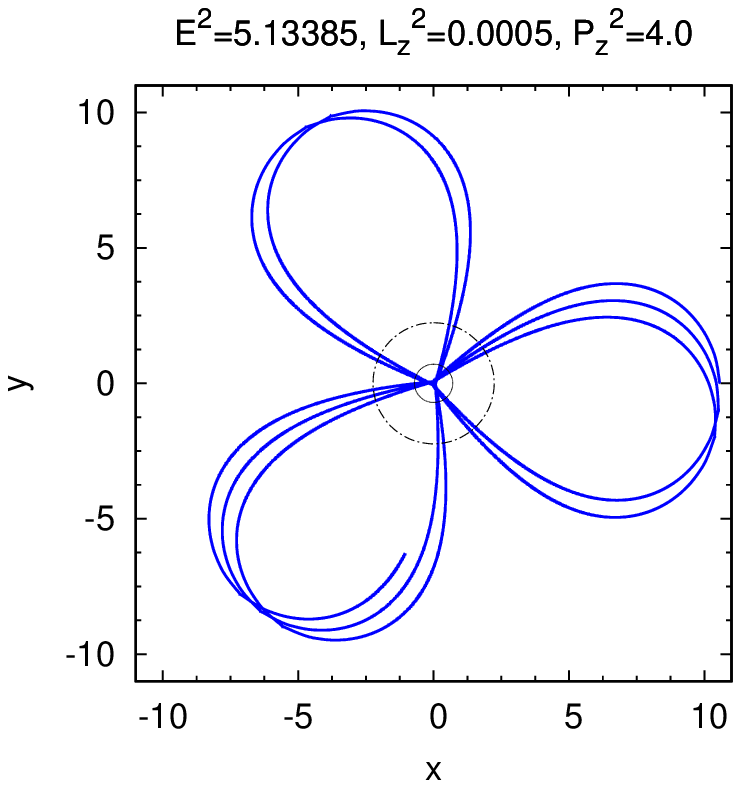}}
\subfigure[][3-d bound orbit]{\label{c7_pot2_orbE2_5.13b_3d}\includegraphics
[width=9cm]{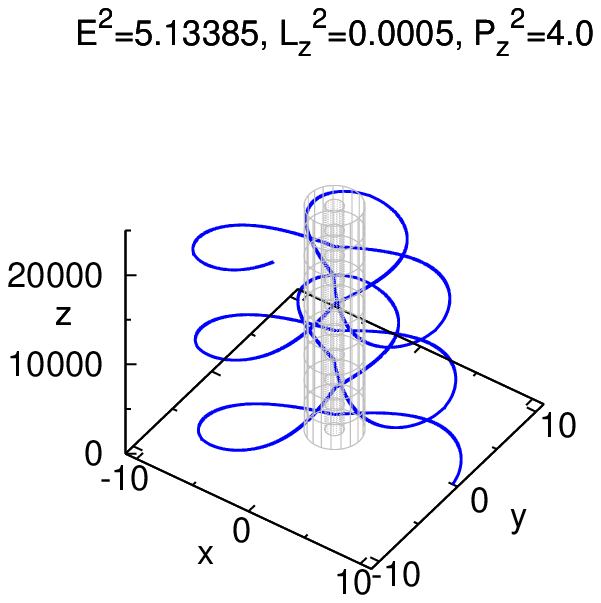}} \\
\subfigure[][escape orbit]{\label{c7_pot2_orbE2_5.14e}\includegraphics
[width=7cm]{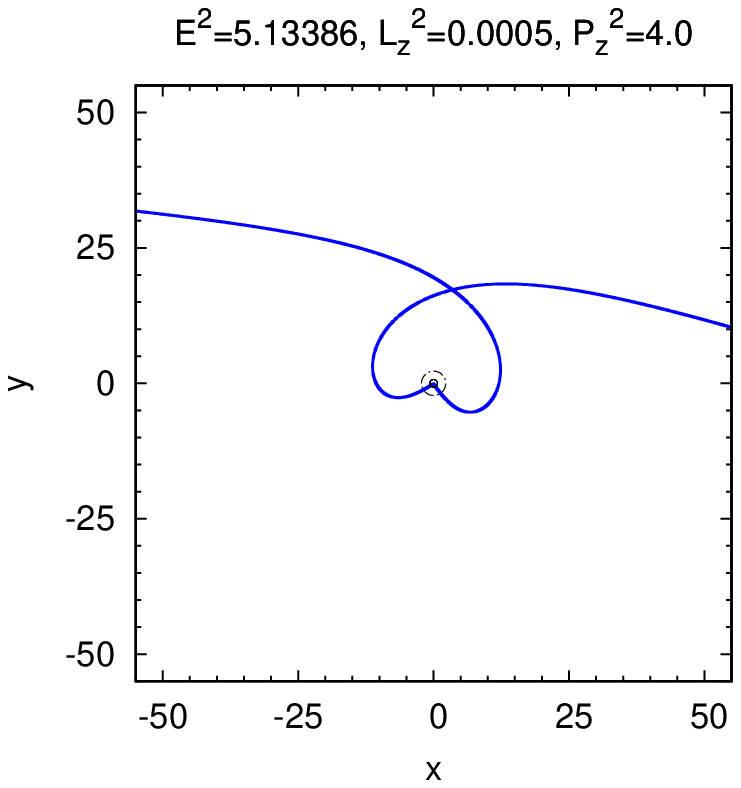}}
\subfigure[][3-d escape]{\label{c7_pot2_orbE2_5.14e_3d}
\includegraphics[width=9cm]{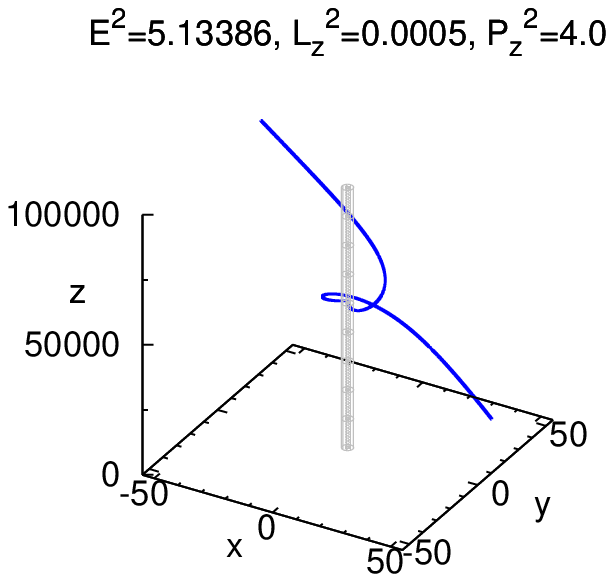}} 
\end{center}
\caption{We show a bound orbit in the $x$-$y$-plane (a) and in the 3-d (b), respectively, as well as an escape orbit
in the $x$-$y$-plane (c) and in 3-d (d), respectively of a massive test particle ($\varepsilon=1$) with 
$L_z^2=0.0005$ and $P_z^2=4$ in the space-time of two Abelian-Higgs strings
interacting via their magnetic fields with $n=1, m=-1, \beta_1=\beta_2=0.2, \alpha=0.001, g=1.0, q=1.0$.
For the bound orbit we have $E^2=5.13385$ and for the escape orbit $E^2=5.13386$, respectively.
The dashed circle and grey cylinder, respectively indicate the string core\label{c7pots_orbits3}.}
\end{figure*}

\begin{figure*}[t]
\begin{center}
\subfigure[][escape orbit]{\label{c7_pot2_orbE2_5.1400e}\includegraphics[width=7cm]{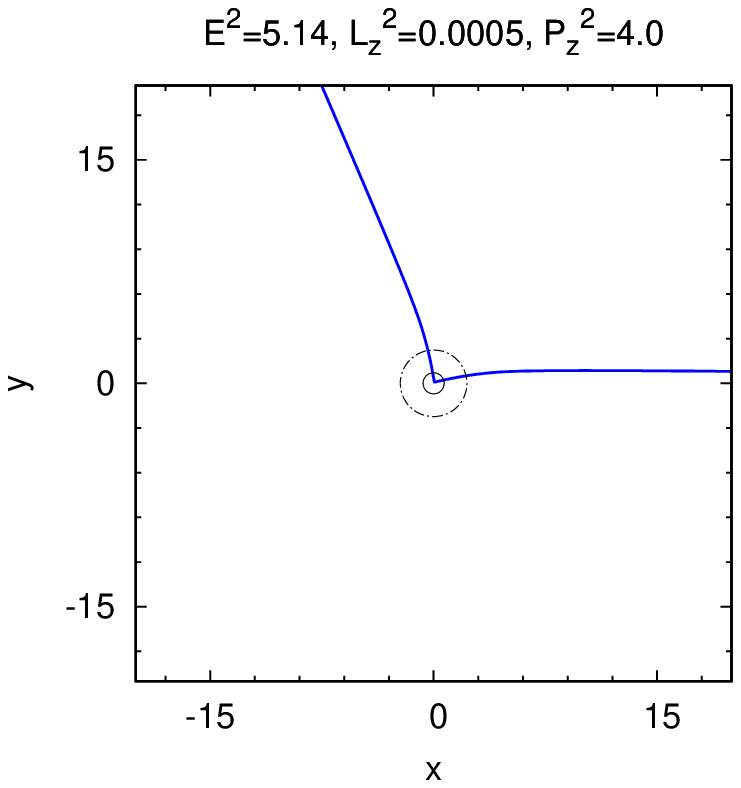}}
\subfigure[][3-d escape orbit]{\label{c7_pot2_orbE2_5.1400e_3d}\includegraphics[width=9cm]
{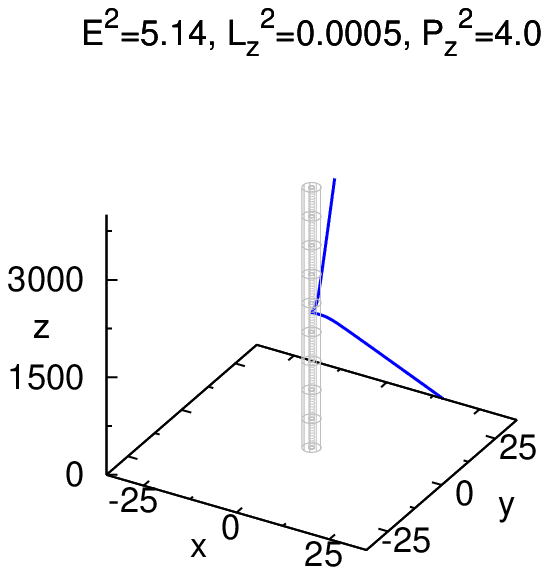}} 
\end{center}
\caption{
We show an escape orbit
in the $x$-$y$-plane (a) and in 3-d (b), respectively of a massive test particle ($\varepsilon=1$) with 
$E^2=5.14$, $L_z^2=0.0005$ and $P_z^2=4$ in the space-time of two Abelian-Higgs strings
interacting via their magnetic fields with $n=1, m=-1, \beta_1=\beta_2=0.2, \alpha=0.001, g=1.0, q=1.0$. 
The dashed circle and grey cylinder, respectively indicate the string core.
\label{c7pots_orbits3_2}}
\end{figure*}

Since the case of negative windings has not been discussed in the literature yet, we also consider this
here. As mentioned above, the effective potential looks qualitatively similar when letting $n$ or $m$ be negative.

In Fig.\ref{c7pots_orbits}-Fig.\ref{c7pots_orbits3_2} we show the orbits of test particles
with $L_z^2=0.0005$, $P_z^2=4$ and different values of $E^2$ in the space-time of two Abelian-Higgs strings with
$n=1$, $m=-1$ $\beta_1=\beta_2=0.2$ and $\alpha=0.001$. 
For small energies (here $E^2=5.01$) we find that the bound orbit lies completely inside the core
of the string (see Fig.\ref{c7pots_orbits} and Fig.\ref{c7pots_orbits2}). Considering
more than one particle this would correspond to a flux of test particles inside the string core.
The maximal radius of the bound orbit
increases with increasing $E$ such that for sufficiently large $E$ the particle moves mainly in the vacuum exterior
region of the string. This is clearly see in Fig.\ref{c7_pot2_orbE2_5.13b} and Fig.\ref{c7_pot2_orbE2_5.13b_3d} for
$E^2=5.13385$. Moreover, the increase in energy has also an effect on the escape orbits. While for small energy
the particle encircles the string (see Fig.\ref{c7_pot2_orbE2_5.14e} and Fig.\ref{c7_pot2_orbE2_5.14e_3d}) it simply
gets deflected by the string for higher values of $E$ (see Fig.\ref{c7pots_orbits3_2}).

\begin{figure*}[t]
\begin{center}
\subfigure[][bound orbit]{\label{c4_pot2_orbE2_5.01b}
\includegraphics[width=7cm]{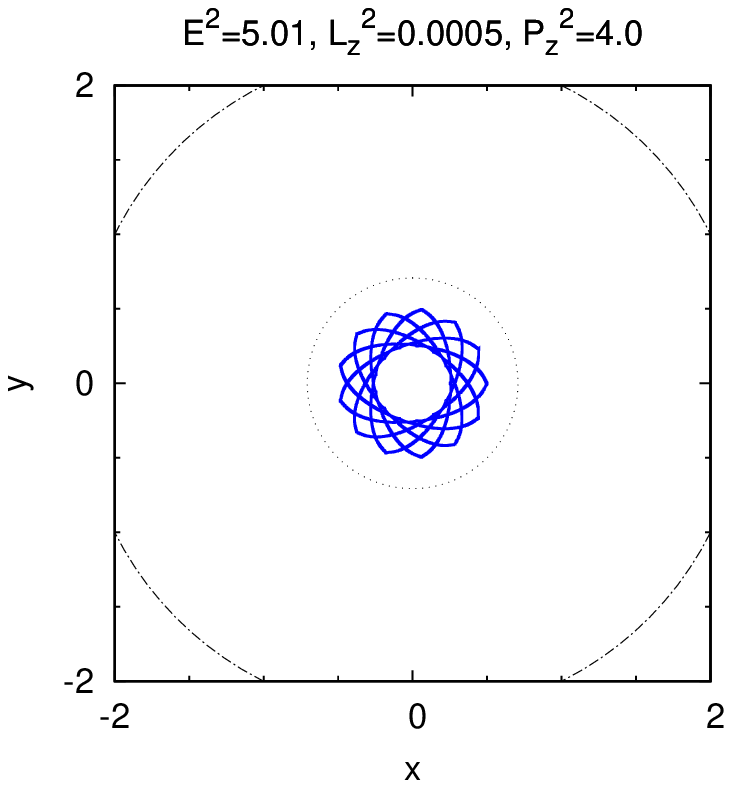}}
\subfigure[][3-d bound orbit]{\label{c4_pot2_orbE2_5.01b_3d}
\includegraphics[width=9cm]{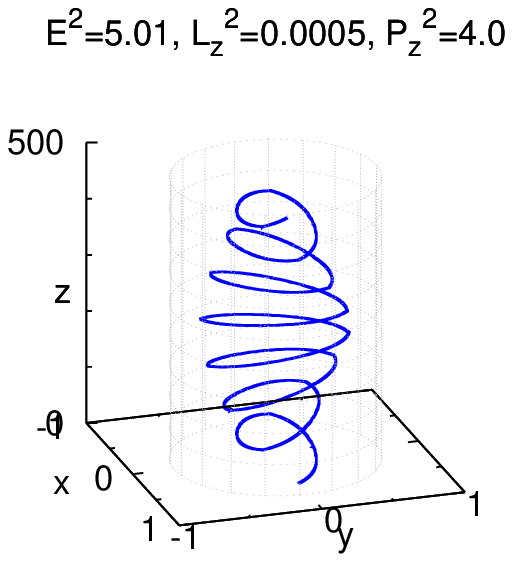}} 
\end{center}
\caption{We show a bound orbit
in the $x$-$y$-plane (a) and in 3-d (b), respectively of a massive test particle ($\varepsilon=1$) with 
$E^2=5.01$, $L_z^2=0.0005$ and $P_z^2=4$ in the space-time of two Abelian-Higgs strings
interacting via their magnetic fields with $n=m=1, \beta_1=\beta_2=0.2, \alpha=0.001, g=1.0, q=1.0$. 
The dashed circle and grey cylinder, respectively indicate the string core.
\label{c4_Lz2_0.0005_orbits}}
\end{figure*}

In order to strengthen the claim that the signature of the windings doesn't have an influence on the
qualitative behaviour of the particles, we plot the bound orbit for $L_z^2=0.0005$, $P_z^2=4$, $E^2=5.01$ 
in the space-time of two Abelian-Higgs strings with
$n=1$, $m=1$ $\beta_1=\beta_2=0.2$ and $\alpha=0.001$ in Fig.s \ref{c4_pot2_orbE2_5.01b}-
\ref{c4_pot2_orbE2_5.01b_3d}. Comparing this with Fig.\ref{c7pots_orbits} we find that the
direction of the magnetic flux (which is given by the choice of signature of the windings) doesn't influence
the motion strongly. The only difference is that the perihelion shift seems to be smaller
for both windings positive.

\subsection{Observables}
The perihelion shift and light deflection of massive and massless test particles, respectively,
has been studied in the space-time of field theoretical cosmic string solutions
previously \cite{hartmann_sirimachan,hartmann_laemmerzahl_sirimachan}. Since our model
is similar, we believe that the qualitative results for these
two observables will be comparable. In this present paper, we hence concentrate
on the computation of the minimal radius of escape orbits of massless test particles
as well as the minimal and maximal radius of bound orbits.
The former has application in the detection of cosmic strings by light deflection, while
in the latter we argue that the radii of bound orbits are on the order of the
inverse gauge boson mass.

The first thing to note is that the minimal and maximal radius of the bound orbits are close to unity (in rescaled
variables) (see e.g. Fig.\ref{c1_pot2_orbE2_5.000134b}).
Reinstalling units, we find that the minimal and maximal radius, respectively are on the order
of the inverse gauge boson mass $M_{W,1}=e_1 \eta_1$. If we assume that 
$M_{W,1} \gtrsim 100 GeV$ we find that one unit of rescaled $\rho$ corresponds
to length scales of $\lesssim 10^{-18} m$. Hence, the orbits
have extremely small extend and are thus not interesting for applications in the context of
motion of massive objects such as planets in the solar system or beyond. However, the fact that
cosmic strings can trap massive test particles that move close to or inside their core might
have interesting cosmological applications (see discussion below).

\begin{figure*}[t]
\begin{center}
\subfigure[][minimal and maximal radius of bound orbit]{\label{rmin_rmax}
\includegraphics[width=8cm]{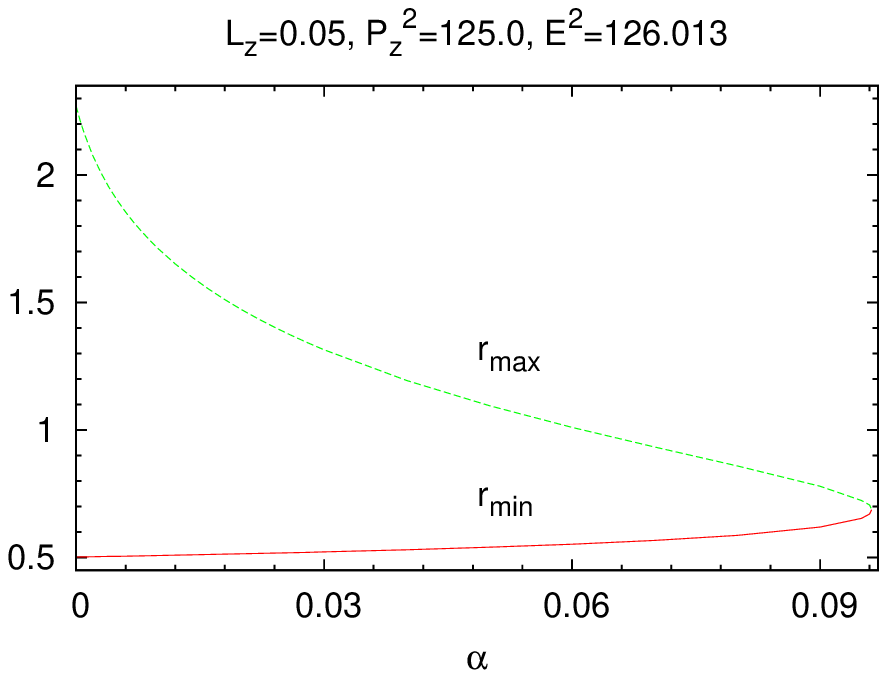}}
\subfigure[][minimal radius of escape orbit]{\label{rmin_escape}
\includegraphics[width=8cm]{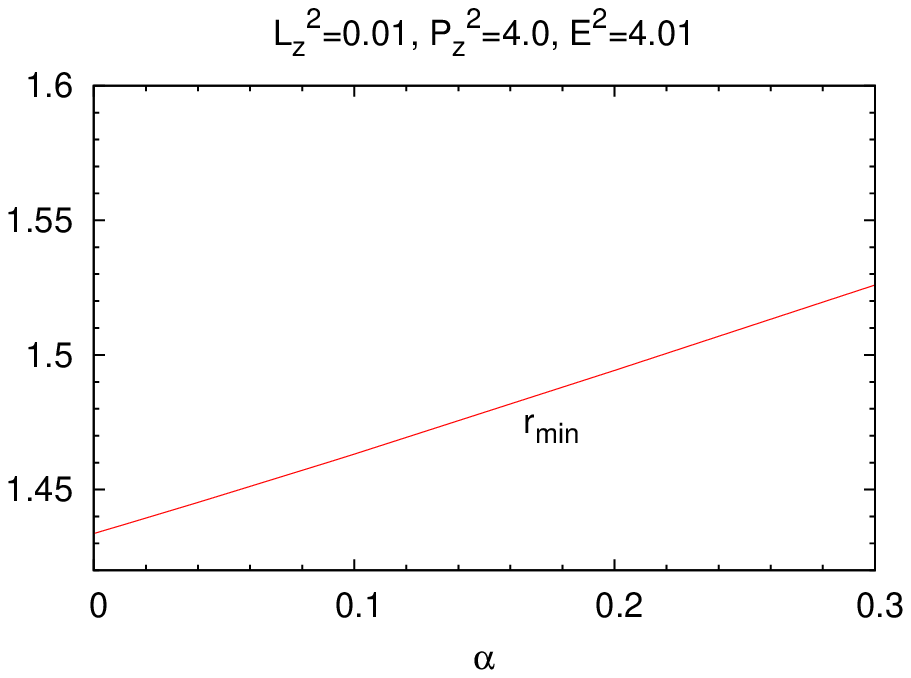}} 
\end{center}
\caption{We show the minimal radius of an escape orbit of a 
massless test particle (left) as well as the minimal and maximal radius of a bound orbit
of a massive test particle (right).
\label{observables}}
\end{figure*}

Massless test particles such as photons can only move on escape orbits, i.e. they get deflected
by the string. While the cosmic strings have very small width in comparison to their lengths and one would
at first think that only the global deficit angle governs the deflection of light, it was
shown already in \cite{hartmann_sirimachan,hartmann_laemmerzahl_sirimachan} that the microscopical
structure of the field configuration has an influence on the observables. It was e.g. found that
massless particles do not simply get deflected by the string, but can encircle it before moving off
again to infinity. This does not happen for infinitely thin cosmic strings. 
In Fig.\ref{rmin_escape} we show the radius of closest approach $r_{\rm min}$ of a massless
test particle moving on an escape orbit in dependence on the interaction parameter $\alpha$.
The radius of closest approach is again on the order of unity in rescaled variables, i.e.
corresponds to length scales of approximately $10^{-18} m$. We observe that the smaller $\alpha$ the smaller
is $r_{\rm min}$, i.e. the deeper the test particle can penetrate into the string core. 

The question is then whether there are any cosmological or astrophysical phenomena observable from these
interactions of massive and massless test particles with cosmic strings. As mentioned above, the most obvious
observation that has already been discussed extensively before is the deflection of light.
Here we want to point out that the motion of massive and massless test particles can lead to the emission
of radiation, in particular gravitational radiation \footnote{We thank Patrick Peter for pointing this out.}.
The emission of scalar, electromagnetic and gravitational radiation, respectively from test particles
moving in the space-time of an infinitely thin cosmic string was first discussed in \cite{aliev_galtsov}.
It was found that the emitted gravitational radiation power $P_{\rm GW}$ of a non-relativistic point particle
moving in the space-time of an infinitely thin cosmic string is given by
\begin{equation}
 P^{\rm(nr)}_{\rm GW} = \frac{\cal E_{\rm GW}}{t_{\rm inter}} = \frac{8\pi/(5r_{\rm min}) G M^2 \beta^2 (v/c)^5}{t_{\rm inter}} \ ,
\end{equation}
where ${\cal E_{\rm GW}}$ is the emitted energy, $t_{\rm inter}$ the interaction time of the particle
with the string, $M$ the mass
of the test particle and $v$ its velocity. Finally $\beta=4G\mu/c^2$ with $\mu$ the energy per unit length of the string.
Now assuming that we have a massive test particle moving on a nearly circular orbit with radius 
$r_{\rm min}$
and hence $t_{\rm inter}\approx 2\pi r_{\rm min}/v$ for one revolution of the particle around the
string, we find the emitted power per revolution:
\begin{equation}
 \tilde{P}^{\rm(nr)}_{\rm GW} = \frac{4}{5 r_{\rm min}^2} G M^2 \beta^2 \left(\frac{v}{c}\right)^5 v \ 
\end{equation}
and $P^{\rm(nr)}_{\rm GW}=\tilde{P}^{\rm(nr)}_{\rm GW}\cdot N$, where $N\in \mathbb{N}$ is the number
of revolutions around the string.

Now assuming that $\beta\approx 10^{-6}$ and $r_{\rm min}\approx 10^{-18} m$ we find 
\begin{equation}
 \tilde{P}^{\rm(nr)}_{\rm GW,1}\approx 5 \cdot 10^{13} M^2 \left(\frac{v}{c}\right)^5 v \frac{m}{kg \ sec^2} \ .
\end{equation}
For non-relativistic point-like, i.e. elementary particle this number is quite small, however,
since the particle moves on a bound orbit it will encircle the string many times and hence the
emitted gravitational radiation can become significant if $N$ becomes very large.

In \cite{aliev_galtsov} the formula for a relativistic point particle was also given. This reads
\begin{equation}
 P^{\rm(r)}_{\rm GW}=\frac{27\pi^2/(32r_{\rm min}) G M^2 \beta^2 \gamma^3}{t_{\rm inter}}
\end{equation}
with the Lorentz factor $\gamma=\left(1-(v/c)^2\right)^{-1/2}$.
Again assuming the particle to move on a bound orbit we find the emitted power per revolution to be
\begin{equation}
 \tilde{P}^{\rm(r)}_{\rm GW}\approx 10^{14} M^2 \gamma^3 v \frac{m}{kg \ sec^2} \ .
\end{equation}
Considering e.g. an electron with $M_{\rm electron}\approx 10^{-31} kg$ moving close to the
speed of light, i.e. $v\lesssim c$ we find that
\begin{equation}
 \tilde{P}^{\rm(r)}_{\rm GW}\approx 10^{-40} \gamma^3 Watt \ .
\end{equation}
$\gamma$ should be very large to have a significant effect here, however, since the
particle is moving in  a bound orbit around the string, the total emitted power could
be quite large.

\section{Conclusions}
Since the advent of inflationary models rooted in String Theory such as brane inflation
it has been suggested that cosmic strings are indeed a ``by-produced'' of inflation
and should hence exist in the universe. 
While the presence of cosmic strings in our universe would show up in the CMB data (power and polarization spectrum)
it would be very exciting indeed to observe these objects directly.  
Since the width of these topological defects is much smaller than their extension it is often
assumed that they are effectively 1-dimensional. Consequently, the Nambu-Goto action
can be used to describe these objects in a so-called ``macroscopic description''. The
advantage is that calculations related to the dynamics of these objects are feasible, however
one doesn't get inside into the underlying field theoretical models. Using the latter in the
so-called ``microscopic description'' however has the disadvantage that even for the simplest
models the solutions have to be constructed numerically. 
There have been claims that certain gravitational lensing effects might be due to cosmic strings
\cite{csl1}. This however turned out to be simply a pair of nearly-identical 
objects and as such cosmic strings are as yet to be detected. We suggest in this paper
that this can be done by the observation of the motion of massive test particles
close to the string core or by the observation of gravitational
lensing, i.e. the motion of massless particles in the gravitational field of a cosmic string.
While in the microscopic description of cosmic strings the space-time is locally flat with 
a global deficit angle and hence geodesics are just straight lines, this is different
for the microscopic description. In that case, bound orbits of massive test particles are possible
\cite{hartmann_sirimachan,hartmann_laemmerzahl_sirimachan}. This is also what we show here.
The cosmic strings interact via their magnetic fields and we show that the attractive interaction
allows for bound orbits that are not possible without the interaction. This was also observed in 
\cite{hartmann_laemmerzahl_sirimachan}. Our model has the new feature that the bound states have a
BPS limit in which they satisfy an equality between their energy per unit lengths and their winding numbers.
BPS states of interacting cosmic strings are of great interest with respect to the 
original supersymmetric p-q-strings appearing in String Theory. 

Since our test particles are point-like, i.e. have no internal structure they interact solely with the
  gravitational field (and hence move on geodesics). It would be interesting to see what would
happen to charged particles or particles with spin. Certainly there will be an effect related
to the interaction with the magnetic field along the cosmic string axis. 
One might consider to compute the electromagnetic and gravitational radiation from this which would
put strong constraints on the energy per unit length of the cosmic strings. This is currently under investigation.

\section*{Acknowledgement} B.H. thanks the Deutsche Forschungsgemeinschaft 
(DFG) for financial support under grant HA-4426/5-1.
V.K. thanks the DFG for financial support. We also 
gratefully acknowledge support within the framework of the DFG Research
Training Group 1620 {\it Models of gravity}. BH would like to thank F. Michel for bringing
reference \cite{aliev_galtsov} to her attention.

\begin{appendix}

\section{No bound orbits for massless test particles}
In \cite{gibbons} it was shown that in a general cosmic string space-time with topology $\mathbb{R}^2\times \Sigma$
bound orbits of massless test particles cannot exist. The assumption used in the proof is that
$\Sigma$ has positive Gaussian curvature. However, it was shown in \cite{hartmann_sirimachan} that
in the space-time of an Abelian-Higgs string $\Sigma$ can have negative Gaussian curvature close to the string
axis and hence the theorem does not apply there. Here we show that for our model bound orbits
of massless test particles are indeed not possible.
In order to have bound orbits we need (at least) two turning points with $d\rho/d\tau=0$. Since
the effective potential $V_{\rm eff}(\rho=0)=+\infty$ for $L_z\neq 0$ (we will discuss the case $L_z=0$ separately)
and $V_{\rm eff}(\rho\rightarrow\infty)\rightarrow c_1^2 \varepsilon +P_z^2 < +\infty$, we need to require
that $V_{\rm eff}$ has local extrema in order to be able to find (at least) two turning points.
Hence $dV_{\rm eff}/d\rho=0$ for some $\rho=\rho_0\neq 0$. For $\varepsilon=0$ and $L_z^2\neq 0$ this leads to
\begin{equation}
 LNN' {\overset{!}=} N^2 L'   \ .
\end{equation}
Taking the derivative and using (\ref{eq6}) and (\ref{eq7}) we find for $\rho\neq 0$
\begin{equation}
\label{condition}
 2 h^2 P^2 + 2 R^2 f^2 + 
\left(\frac{P'}{g} - R'\right)^2 + R'^2 +\frac{2}{g} (1-\alpha)  R' P' {\overset{!}=} 0 \ .
\end{equation}
Now, since $R'< 0$, $P'<0$ (both functions monotonically decrease for $n$, $m$ at $\rho=0$ to
zero at infinity, see (\ref{eq4}), (\ref{eq5})) we find that for $0 \leq \alpha \leq 1$ (\ref{condition}) can never
be fulfilled. For $\varepsilon=0$ and $L_z^2=0$ the potential is constant $V_{\rm eff}(\rho)\equiv P_z^2$.
Hence the potential cannot have local extrema and we conclude that bound orbits of massless test particles are not possible. 
Note that our model includes also the case of ordinary Abelian-Higgs strings (for $\alpha=0$) that was
previously studied in \cite{hartmann_sirimachan} as well as the case of
Abelian-Higgs strings interacting via their potential that has
been considered in \cite{hartmann_laemmerzahl_sirimachan}. The latter applies because the field potential (\ref{potential_scaled})
drops out when combining (\ref{eq6}) and (\ref{eq7}). Hence, even if direct interaction terms between the scalar fields
would be considered in the potential they wouldn't effect the argument above.
 
\end{appendix}

\bibliographystyle{unsrt}

\begin{thebibliography}{}

\end{thebibliography}


\begin{thebibliography}{99} 

\bibitem{kibble} T. Kibble, J. Phys. A {\bf 9} 1378 (1976).
\bibitem{no} 
 H.~B.~Nielsen and P.~Olesen,
  Nucl.\ Phys.\  B {\bf 61}, 45 (1973).
\bibitem{vs} A. Vilenkin and  P. Shellard, {\it Cosmic strings and other topological defects}, Cambridge University Press (1994).
\bibitem{polchinski} see e.g. J. Polchinski, {\it Introduction to cosmic
F- and D-strings}, hep-th/0412244 and reference therein.

\bibitem{braneinflation} 
M.~Majumdar and A.~C.~Davis,
  JHEP {\bf 0203} (2002) 056
  [arXiv:hep-th/0202148].
S.~Sarangi and S.~H.~H.~Tye,
  Phys.\ Lett.\  B {\bf 536}, 185 (2002)
  [arXiv:hep-th/0204074].



\bibitem{lyth}
  D.H. Lyth and A. Riotto,  Phys.\ Rept.\  {\bf 314} (1999) 1.

\bibitem{jeannerot}
  R. Jeannerot, J. Rocher and M. Sakellariadou,  Phys.\ Rev.\  D {\bf 68} (2003) 103514
  

\bibitem{saffin}
  P.M. Saffin,  JHEP {\bf 0509} (2005) 011.

\bibitem{rajantie} A. Rajantie, M. Sakellariadou and H. Stoica, JCAP {\bf 11} 021 (2007). 
\bibitem{salmi} P.Salmi {it et al},  Phys.\ Rev.\  D {\bf 77} 041701 (2008).
\bibitem{urrestilla}
  J. Urrestilla and A. Vilenkin,
  JHEP {\bf 0802} 037 (2008).



\bibitem{hartmann_urrestilla} B. Hartmann and J. Urrestilla, JHEP {\bf 07} 006 (2008). 

\bibitem{clv} M. Christensen, A.L. Larsen and Y. Verbin, Phys. Rev. D {\bf 60}, 125012 (1999).
\bibitem{bl} Y. Brihaye and M. Lubo, Phys. Rev. D {\bf 62}, 085004 (2000).
\bibitem{ha}  B.~Hartmann and F.~Arbabzadah,
  JHEP {\bf 0907} (2009) 068.
\bibitem{bps} E.~B.~Bogomolny,
  Sov.\ J.\ Nucl.\ Phys.\  {\bf 24} (1976) 449
  [Yad.\ Fiz.\  {\bf 24} (1976) 861].

\bibitem{cmb}
N. Bevis {\it et al},  Phys. Rev. {\bf D75}, 065015 (2007);  
N. Bevis {\it et al},  arXiv:astro-ph/0702223;
N. Bevis {\it et al},  Phys. Rev. {\bf D76}, 043005 (2007);
N.~Bevis, M.~Hindmarsh, M.~Kunz and J.~Urrestilla,
  Phys.\ Rev.\ Lett.\  {\bf 100}, 021301 (2008);
 Phys.\ Rev.\  D {\bf 75}, 065015 (2007);
Phys.\ Rev.\ D {\bf 82}, 065004 (2010);
JCAP {\bf 1112}, 021 (2011).


\bibitem{cmb2}
  J.~Urrestilla, P.~Mukherjee, A.~R.~Liddle, N.~Bevis, M.~Hindmarsh and M.~Kunz,
  Phys.\ Rev.\ D {\bf 77}, 123005 (2008);
Phys.\ Rev.\ D {\bf 83}, 043003 (2011).

\bibitem{csl1} M.~V.~Sazhin {\it et al.},
  Mon.\ Not.\ Roy.\ Astron.\ Soc.\  {\bf 376} (2007) 1731;
  M.~V.~Sazhin, M.~Capaccioli, G.~Longo, M.~Paolillo and O.~S.~Khovanskaya,
  arXiv:astro-ph/0601494; Astrophys.\ J.\  {\bf 636} (2005) L5.

\bibitem{ag} 
 A.~N.~Aliev and D.~V.~Galtsov,
  Sov.\ Astron.\ Lett.\  {\bf 14}, 48 (1988).
\bibitem{gm} D.~V.~Galtsov and E.~Masar,
  Class.\ Quant.\ Grav.\  {\bf 6}, 1313 (1989).
\bibitem{cb}  S.~Chakraborty and L.~Biswas,
  Class.\ Quant.\ Grav.\  {\bf 13}, 2153 (1996).
\bibitem{Ozdemir2003} N.~Ozdemir,
  Class.\ Quant.\ Grav.\  {\bf 20} 4409 (2003).
\bibitem{Ozdemir2004}
  F.~Ozdemir, N.~Ozdemir and B.~T.~Kaynak,
  Int.\ J.\ Mod.\ Phys.\  A {\bf 19} 1549 (2004).
\bibitem{hhls1} E.~Hackmann, B.~Hartmann, C.~L\"ammerzahl and P.~Sirimachan,
  Phys.\ Rev.\  D {\bf 81}, 064016 (2010)
  [arXiv:0912.2327 [gr-qc]].

\bibitem{hhls2} E.~Hackmann, B.~Hartmann, C.~L\"ammerzahl and P.~Sirimachan,
 Phys.\ Rev.\  D {\bf 82}, 044024 (2010)
  [arXiv:1006.1761 [gr-qc]].

\bibitem{hartmann_sirimachan} B. Hartmann and P. Sirimachan, JHEP {\bf 08} 110 (2010). 


\bibitem{hartmann_laemmerzahl_sirimachan} B. Hartmann, C. L\"ammerzahl and P. Sirimachan, Phys. Rev. D {\bf 83} 045027 (2011). 

\bibitem{dark} N. Arkani-Hamed, D. Finkbeiner, T. Slatyer and N. Weiner, Phys.Rev.D {\bf 79}, 015014 (2009); N. Arkani-Hamed and N. Weiner, JHEP {\bf 0812}, 104 (2008);
L. Bergstrom, G. Bertine, T. Bringmann, J. Edsjo and M. Taoso, arXiv: 0812.3895 [astro-ph].
\bibitem{vachaspati}  T.~Vachaspati,
  Phys.\ Rev.\  D {\bf 80} (2009) 063502
  [arXiv:0902.1764 [hep-ph]].
\bibitem{GK2011} S. Grunau, V.~Kagramanova, Phys. Rev. D {\bf 83} 044009 (2011).


\bibitem{kkl} 
V.~Kagramanova, J.~Kunz and C.~Lammerzahl,
  Gen.\ Rel.\ Grav.\  {\bf 40} 1249 (2008).


\bibitem{colsys} U. Ascher, J. Christiansen and R. Russell, Math. of Comp. {\bf 33}, 659 (1979);
ACM Trans. {\bf 7}, 209 (1981).  


\bibitem{gl} D. Garfinkle and P. Laguna, Phys. Rev. D {\bf 39}, 1552 (1989);
M. E. Ortiz, Phys. Rev. D {\bf 43}, 2521 (1991).


\bibitem{gibbons}  G.~W.~Gibbons,
  Phys.\ Lett.\  B {\bf 308}, 237 (1993).

\bibitem{aliev_galtsov} A.~N.~Aliev and D.~V.~Galtsov,
  Annals Phys.\  {\bf 193}, 142 (1989).



\end{thebibliography}

\end{document}